%% file: main.tex
\documentclass[trackchanges,twocolumn]{aastex701}

\newcommand{\unit}[1]{\ensuremath{\,\mathrm{#1}}}
\newcommand{\um}{\unit{\mu m}}
\newcommand{\km}{\unit{km}}
\newcommand{\au}{\unit{au}}
\newcommand{\pct}{\unit{\%}}

\newcommand{\spherex}{SPHEREx}
\newcommand{\atlas}{3I}
\newcommand{\hartley}{103P}

\newcommand{\water}{\ensuremath{\mathrm{H_2 O}}}
\newcommand{\co}{\ensuremath{\mathrm{CO}}}
\newcommand{\range}[2]{\ensuremath{#1 \mathrm{-} #2}}

\usepackage[x11names,dvipsnames]{xcolor}

\begin{document}

\title{SPHEREx Pre-Perihelion Mapping of $\mathrm{H_2O}$, $\mathrm{CO_2}$, and $\mathrm{CO}$ in Interstellar Object 3I/ATLAS
}

\input{author_list_3i-atlas_2025aug.tex}

\begin{abstract}
From 01- to 15-Aug-2025UT, the \spherex\ spacecraft observed interstellar object 3I/ATLAS. Using $R=\range{40}{130}$ spectrophotometry at $\lambda=\range{0.7}{5}\um$, light curves, spectra, and imaging of \atlas\ were obtained. From these, robust detections of water gas emission at $\range{2.7}{2.8}\um$ and $\co_2$ gas at $\range{4.23}{4.27}\um$ plus tentative detections of $^{13}\co_2$ and $\co$ gas were found. A slightly extended $\water$ coma was detected, and a huge $\co_2$ atmosphere of extending out to at least $4.2\times10^{5}\km$ was discovered. Gas production rates and 1$\sigma$ errors for $\water$, $^{12}\co_2$, $^{13}\co_2$, and $\co$ were $ Q_{gas} = 3.2 \times 10^{26} \pm 20 \pct $, $ 1.6 \times 10^{27} \pm 10 \pct$,  $1.3 \times 10^{25} \pm 25 \pct$, and  $1.0 \times 10^{26} \pm 25 \pct$, respectively. Co-addition of all $\lambda = \range{1.0}{1.5} \um$ scattered light continuum images from produced a high SNR image consistent with an unresolved source. The scattered light lightcurve showed $\lesssim 15 \pct$ variability  over the observation period. The absolute brightness of \atlas\ at $\range{1.0}{1.5}\um$ is consistent with a $< 2.5 \unit{km}$ radius nucleus surrounded by a 100 times brighter coma. The $\range{1.5}{4.0}\um$ continuum structure shows a strong feature commensurate with water ice absorption seen in KBOs and distant comets. The observed cometary behavior of \atlas, including its preponderance of $\co_2$ emission, lack of $\co$ output, small size, and predominance of large icy chunks of material in a flux-dominant coma is similar to the behavior of short period comet 103P/Hartley 2, the ''hyperactive comet'' flyby target of the NASA Deep Impact extended mission in 2010. \textcolor{black}{This correspondence suggests that ISOs can be significantly thermally processed before ejection into the ISM, and by comparison to 1I and 2I, can be widely variable in their physical outcome.} 
\end{abstract}


\section{Introduction} \label{s:intro}

\spherex\ (Spectro-Photometer for the History of the Universe, Epoch of Reionization and Ices Explorer) is a new NASA mission launched into low-Earth, Sun-synchronous orbit (altitude $\approx 675 \km$) on 11-Mar-2025 UT \citep{2016arXiv160607039D, 2018arXiv180505489D, 2025arXiv251102985B}. Conducting a 102-band near-infrared spectrophotometric survey of the entire sky over the course of 2 years, it was designed primarily to study (1) inflationary cosmology, (2) the history of galaxy formation, and (3) the abundance of astrobiologically important molecular ices in planet-forming regions.  \spherex\ will observe and catalog measurements of everything along a line of sight on the sky---from local solar system objects (asteroids, comets, planets, KBOs) to main sequence stars, to distant quasars. One of these objects, 3I/ATLAS (\atlas, hereafter), was recently discovered on 01-Jul-2025 \citep{2025MPEC....N...12D} and verified within the next few days as only the 3rd ever macroscopic body detected hurtling through the solar system from the depths of interstellar space. Termed ``ISOs'', bodies like \atlas\ represent unique samples of exosystem material delivered close to Earth ``for free'', allowing in-depth remote sensing analysis of their size, shape, composition, and temporal behavior during the few months when they are passing through the inner solar system.

Analysis of \atlas's orbit and apparent motion on the sky [JPL Horizons Ref] with respect to the \spherex\ sky survey region (Solar elongation $\sim 90\degr$; \cite{2025arXiv250820332B}) showed that the object would be visited $\sim~70$ times from 08- to 15-Aug-2025 during the normal survey (Fig. \ref{fig:orbit}), with some gaps in wavelength coverage, especially around $3 \um$, due to \atlas's rapid apparent sky motion. The project then proceeded to design, test, and implement a special observing mode, only 2 months after the start of the mission's all-sky survey, to augment the total number of visits to \atlas\ up to 160 flux measurements while completely characterizing the object from $\range{0.75}{5.0} \um$.

The initial results from the first $\sim70$  \spherex\ normal survey visits to \atlas, as published in \cite{2025RNAAS...9..242L}, resulted in the discovery of a giant $\co_2$ coma surrounding the nucleus and the presence of large ($> 100 \um$) water-ice rich grains in the coma. Scattered light was detected from the nucleus and coma grains at $\lambda \lesssim 2.3\um$, but only 3-$\sigma$ upper limits were placed on the emission rates of $\water$ and $\co$ gas.

In this paper, with more than double the total measurements and a much better data reduction and data calibration, we extend the preliminary \spherex\ discoveries to include a robust detection of the water gas emission feature at $\range{2.7}{2.8}\um$ and a marginal detection of $\co$ gas emission at $4.7 \um$.

In addition, the detected $^{12}\co_2$ line emission at $\range{4.25}{4.27}\um$ was so strong that we searched for substructure in it. While we were not able to resolve the $4.25$ vs $4.27 \um$ doublet at the peak, we were able to spectrally distinguish a shoulder at $4.37 \um$ where emission due to $^{13}\co_2$ is found. Assuming the same distribution pattern as for the $^{12}\co_2$ and thus the same aperture size, we find the abundance ratio of $^{13}\co_2$ to $^{12}\co_2$ to be the same, within the error bars of the measurement, as reported by \cite{2025ApJ...991L..43C} using JWST NIRSPEC IFU spectroscopy.

\spherex's unique observational strengths lie in its ability to spectrophotometrically image an object at $\range{0.75}{5.0} \um$ over many weeks and on large angular scales. \spherex's spectrophotometric quality is attested to by its overall excellent \atlas\ spectral match to the published $\range{0.7}{5.3}\um$ JWST spectrum \citep{2025ApJ...991L..43C}; where they differ, in the gas emission lines, it is always \spherex's measurement that is the larger due to its much larger projected observing aperture (Fig. \ref{fig:mjy+refl}). (It is important to note that the area of the \spherex\ central pixel is 4 times the area of the entire JWST NIRSPEC IFU field of view.) \spherex's temporal capabilities are demonstrated by the 12 day long dust continuum reflectance lightcurve (Fig. \ref{fig:orbit}c) and its large field image quality by $\sim 6\arcmin$ wide $\co_2$ coma detection shown in Fig. \ref{fig:stack}.

The chief systematic found for the \spherex\ \atlas\ observations is confusion with background stars. Other effects, like motion with respect to the Sun and Earth, and changes in phase angle, were small and easily corrected for. Object variability, which can introduce large uncertainties in \spherex\ spectrophotometry obtained piecemeal over 15 days, was not an issue due to the lack of any observed lightcurve variation greater than 15\% found (Fig. \ref{fig:orbit}c) and in reports by other groups \citep{2025ApJ...990L..65K, 2025arXiv250713409C}. This lack of variation, in turn, is likely due to the dominance of the observed flux by the slowly time-varying smooth coma, which can mask any rapid projected surface area variations of an asymmetric rotating nucleus.

In this Letter we describe the complete set of August 2025 \spherex\ observations of \atlas, and their data quality, reduction, and analysis. Because of these observations, we now understand that the extended optical morphology seen for \atlas\ since its discovery is the product of robust $\co_2$ gas sublimation that was undetectable in the first UVIS characterizations of the object reported in July 2025.
We also find a very $\co$-poor but $\water$-ice rich cometary object with its water gas production just beginning to turn on, and evidence for thermal processing of the body as seen in highly thermalized inner solar system comets. We find good concordance with other August 2025  imaging and spectroscopic measures of the object. We use these findings to argue that \textcolor{black}{the surface layers} of \atlas\ must have been well-thermally processed before its ejection from its home system into the ISM, and that such thermal processing happens more rapidly than exosystems dynamically relax.

\section{Observations} \label{s:obs}

\begin{figure*}
\centering
   \includegraphics[width=\textwidth]{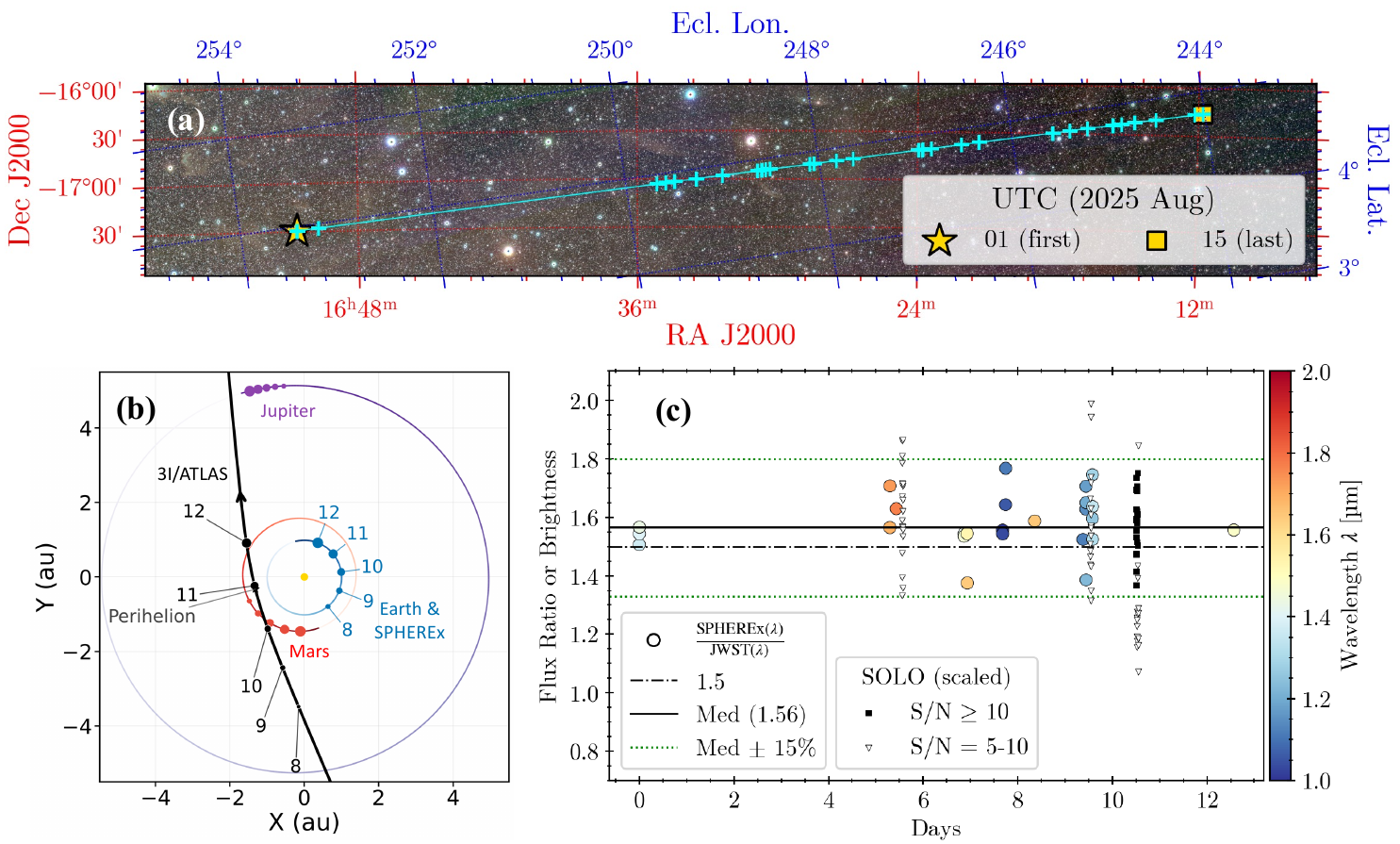}
   \caption{
        Motion of \atlas. \textbf{(a)} Color composite from \spherex\ data at 1.185, 1.716, and 2.194 $\um$ for red, green, and blue, respectively. The trajectory of \atlas\ is overplotted, with observed positions marked by crosses. The starting and ending points of our dataset are indicated separately (see legend). \textbf{(b)} Orbital motion of \atlas\ (black) projected onto the ecliptic plane. The orbits of Earth, Mars, and Jupiter are shown in blue, red, and purple, respectively. Numbers indicate the positions of the bodies at the first day of each month. \atlas's retrograde motion and perihelion are marked with arrows. \textbf{(c)} Lightcurve of \atlas\ produced by taking all $\lambda = \range{1.0}{1.8} \um$ \spherex\ flux measurements and dividing them by the JWST flux (Fig. \ref{fig:mjy+refl}b; \citealt{2025ApJ...991L..43C}) to leave only lightcurve variations. Also included are SOLO clear-filter brightness measurements scaled to match their median to that of the \spherex/JWST ratio. The resulting lightcurve amplitude is within $\pm 15\pct$, so the derived reflectance should be correct to within this range considering any lightcurve modulation effects.
   }
   \label{fig:orbit}
\end{figure*}

\spherex\ obtained 160 frames with \atlas\ within the field of view from 2025-08-01 UT 08:01:07 to 2025-08-15 UT 12:43:48 (midpoints of exposures), including 94 exposures produced by deliberately adjusting the survey planning algorithm to place \atlas\ in the FoV. A subset of these were downlinked and processed promptly, leading to our express publication in \cite{2025RNAAS...9..242L}. In this work, we analyze all 160 frames (Table \ref{tab:obs}). The orbital geometry of the observation and the sky mosaic images are shown in Fig. \ref{fig:orbit}.

Ancillary lightcurve observations were obtained using the remote/robotic Solar system Objects Light-curve Observatory (SOLO; B. Lim et al., submitted to JKAS), a Celestron LLC’s RASA11 telescope\footnote{An 11-inch reflecting telescope manufactured by Celestron LLC} located at Sierra Remote Observatory\footnote{IAU observatory code G80, $37\degr 4\arcmin 13.39\arcsec$ N, $119\degr 24\arcmin 45.72\arcsec$ W}, CA, USA in order to independently check the \spherex\ time domain trending. Combined with the currently installed imaging sensor at the prime focus, KL4040\footnote{Manufactured by FLI Kepler, a 4k by 4k CMOS Back Illuminated (BI) camera}, and a UV-cut clear filter, the total throughput of the SOLO system as of writing is effectively sensitive to $\sim \range{0.4}{1.0} \um$, with a response closely comparable to the Gaia G band. The results are shown in Fig. \ref{fig:orbit}c, and are consistent with the low (if any) level of variability detected by \spherex.

Ancillary spectral observations were used to independently check the \spherex\ spectral measurements were obtained at the NASA Infrared Telescope Facility (IRTF) on the nights of 09 and 10-Aug-2025 using the SpeX instrument in $R\sim50$, $\lambda = \range{0.7}{2.5} \um$ prism spectroscopy mode (Fig. \ref{fig:mjy+refl}). These measurements were vital for verifying initial \spherex\ spectrophotometric data reductions and quickly bootstrapping early photometric measures, and the two sets of measurements were found to be closely consistent, especially the strong dropoff past $2.3\um$ in the IRTF spectrum that was initially believed to be an instrumental artifact near the end of the spectral order.

\section{Data Reduction} \label{s:res}
We first simulated the exposures using the \spherex\ reference catalog (Y. Yang et al., in prep.) and the \spherex~Sky Simulator \citep{2025ApJS..281...10C}. The simulated images, which do not have \atlas\ in them, were then subtracted from the calibrated raw Level 2b (l2b) image data provided  by the \spherex\ science pipeline in Quick Release 2 (QR2; \spherex\ Explanatory Supplement). This was extremely helpful in removing the estimated flux of field objects within \atlas's aperture, flagging less reliable images and extendedness artifacts, and significantly improved the final spectrum.

For signal extraction, we applied simple circular aperture photometry with a $12\arcsec$ radius ($\sim2$ pixels, depending on the location in the FoV due to the distortion). The residual local sky background was estimated from a circular annulus with inner and outer radii of $\approx 1\arcmin$ and $3 \arcmin$, respectively. The photometry at the same celestial location from the original l2b image, simulated image, and the simulation-subtracted image are $F_\mathrm{2b}$, $F_\mathrm{sim}$, and $F_\lambda = F_\mathrm{2b} - F_\mathrm{sim}$, respectively.

We verified that the results are robust against variations in the annulus size and different sky estimation algorithms. As an additional \textit{null test}, we measured the single pixel value minus the estimated sky from the surrounding annulus for several pixels that are neither flagged as bad nor expected to contain source signals (based on the reference catalog) within each image. These values are consistent with zero, with scatter comparable to the expected pixel variance. This confirms that the sky estimation is reliable at the pixel-variance level.

The uncertainty of $F_\lambda$, $dF_\lambda$, is obtained based on the aperture sum of the associated variance map of the l2b image, quadratically combined with the local sky-estimation error. The variance map includes both readout and Poisson noise terms \citep{2025ApJS..281...10C, SPHEREx_Expsupp_QR_v1.3}. Since this does not take into account the uncertainty of the fluxes of field objects, it should be regarded as a lower limit; however, it should be close to the true value when sidereal object contamination within the aperture is small.

We define the contamination factor as
\begin{equation}\label{eq:F_field}
    F_\mathrm{field} \equiv F_\mathrm{sim} / F_{2b} ~.
\end{equation}
For example, if a contaminant contributes half the flux of \atlas, then we expect $F_\mathrm{field} = 0.5/(1+0.5) \approx 0.3$. The higher this number, the larger the measurement uncertainty.

Among the 160 exposures, 9 were flagged as ``a'' and are completely discarded because at least one unrecoverable pixel (cold, hot, non-functional, early transient, and early overflow) is within the aperture, and it is unlikely that they can ever be improved. Flag ``b'' is given to 59, and they are regarded as less reliable as of writing because of strong background source contamination expected and/or calibration ($F_\mathrm{field} > 30$\,\%, object is on the edge of detectors, etc). Finally, 11 were flagged ``c'', which means we see a non-negligible contamination likely from background objects identified by visual inspection, that is, the flux interpolated from the reference catalog does not fully represent the sidereal sources near \atlas. In principle, fluxes from exposures with flags ``b'' and ``c'' could be improved once \spherex\ obtains images of the same sky regions at the same wavelengths, when suitable ``reference'' images will be available for subtraction.

\section{Results} \label{s:res}
The direct observables produced by \spherex\ spectrophotometric imaging of \atlas\ in early August 2025 fall into 3 broad categories: temporal lightcurves, spectrophotometry, and imaging.

\subsection{Lightcurves}
 Lightcurve analysis of \atlas's temporal behavior (Fig. \ref{fig:orbit}) shows very little variability of $\lesssim 15\pct$ from hours to days timescales, which is also confirmed by independent SOLO observations. This upper limit of variation is also consistent with previous reports \citep{2025ApJ...989L..36S, 2025ApJ...990L..65K, 2025PASJ...77L..71B}.  The fortuitous lack of temporal variability, explained by an object with a small, likely dark nucleus surrounded by a very bright and high surface area coma (see Section~\ref{sec_nucleus} below), meant that we could quickly cull out any background contaminated images and proceed straight to analysis of the data.

\subsection{Spectra} \label{ss:spec}
\begin{figure*}
\centering
\plotone{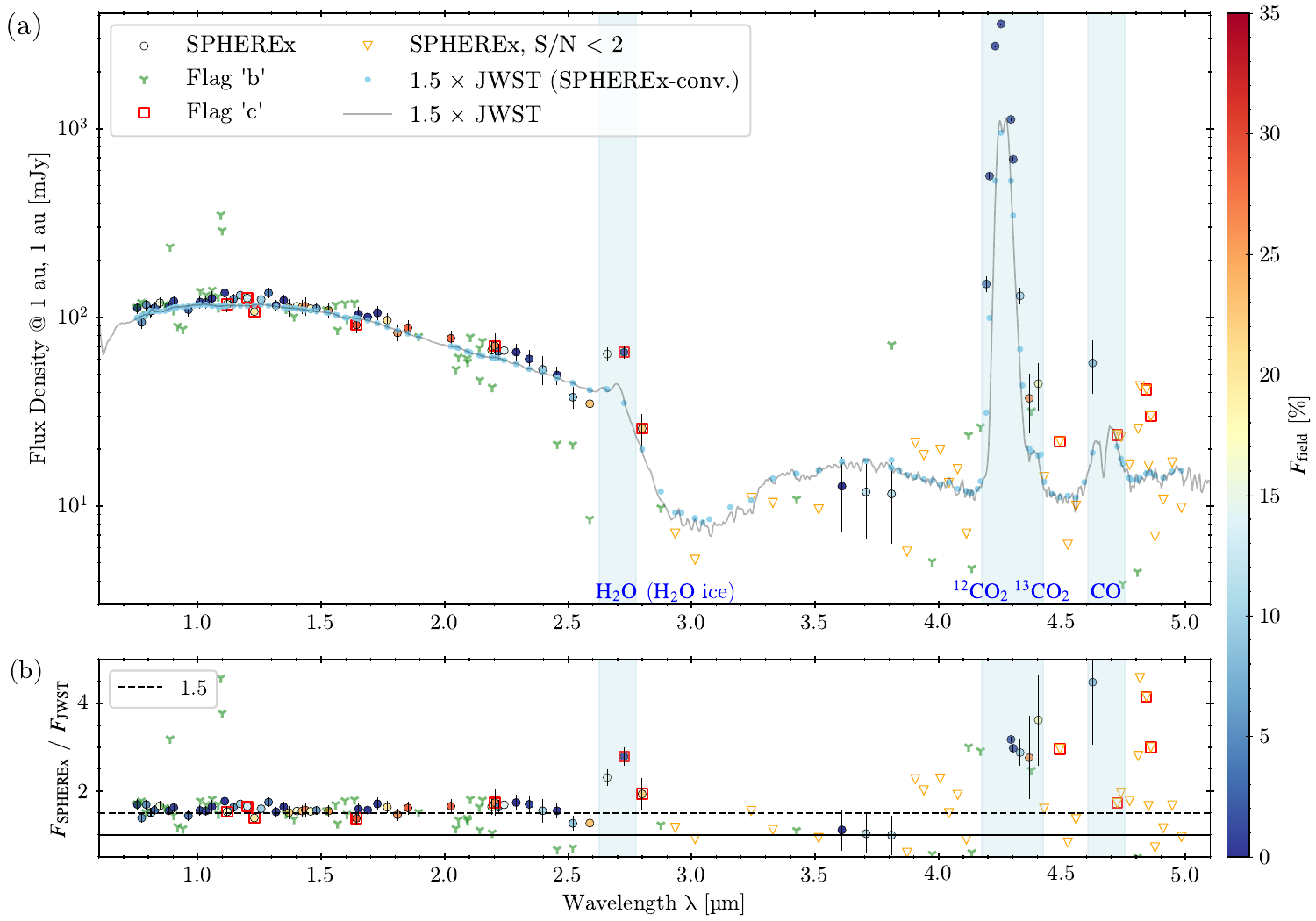}
\plotone{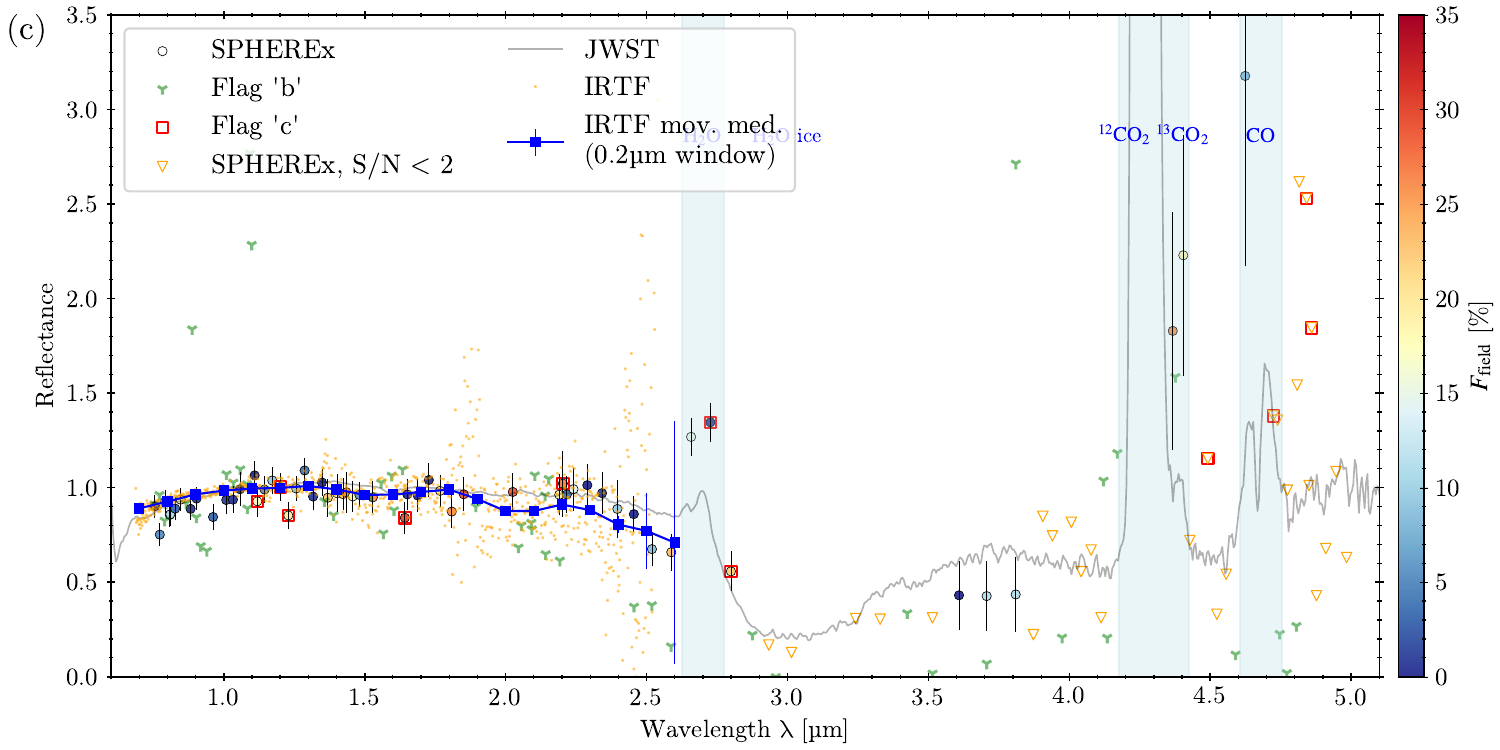}
   \caption{
        \textbf{(a)} The flux density of the measurements, scaled to the geometry of $r_\mathrm{obs} = r_\mathrm{hel} = 1 \,\mathrm{au}$. The JWST measurements \citep{2025ApJ...991L..43C} are scaled by a constant factor of 1.5 to match the \spherex\ measurements. For flags, see Table \ref{tab:obs} and text for explanation. The skyblue points are the JWST measurements, convolved with the pixel spectral throughput at the \spherex\ pixel for that observation, using the \spherex\ Sky Simulator \citep{2024SPIE13092E..3NH, 2025ApJS..281...10C}. Blue shades indicate important spectral features. Colors of ``good'' measurements indicate $F_\mathrm{field}$ (Eq. \ref{eq:F_field}). \textbf{(b)} The distance-corrected ratio of the \spherex\ to JWST flux, showing that \spherex\ measurements are systematically higher than JWST. This indicates the spatial extendedness of the object, \atlas. The error-bars are not shown for $S/N < 2$ data.
        \textbf{(c)} The reflectance, normalized at $1.2\um$ for \spherex, JWST, and the IRTF measurements. The blue line shows the moving median of the IRTF data with a $0.2\um$ window.
   }
   \label{fig:mjy+refl}
\end{figure*}

The measured fluxes of \atlas\ are scaled to the geometry of $r_\mathrm{obs} = r_\mathrm{hel} = 1 \,\mathrm{au}$ assuming $F_\lambda \propto r_\mathrm{hel}^{-2} r_\mathrm{obs}^{-2}$ (Fig. \ref{fig:mjy+refl}a). The reflectance spectrum is derived by dividing the \spherex\ measurements by the solar flux. We can break up the spectrum into 5 main regions: the $\range{0.7}{1.5} \um$ scattered light region; the $\range{1.5}{2.3} \um$ mixed organics plus water absorption region; the $\range{2.4}{3.3} \um$ water ice absorption region; the $\range{3.2}{3.7} \um $ organics absorption region; and the $\range{3.7}{5.0} \um$ mixed region. Superimposed on top of the continua are weak gas emission features due to $\water$ at $\range{2.7}{2.8} \um$ and $\co$ at $\range{4.6}{4.8} \um$, and a strong $\co_2$ emission feature at $\range{4.2}{4.4} \um$. The $\co_2$ feature is strong enough that we believe we have detected not only normal $^{12}\co_2$ centered in a doublet at $4.25$ and $4.27 \um$, but also a small side shoulder due to $^{13}\co_2$ centered at $4.31 \um$. The agreement in spectral shape with the JWST spectrum of \cite{2025ApJ...991L..43C} taken in one visit during the middle of our observational window is excellent; the main disagreement between the two spectral datasets is in the amplitude of the gas emission lines, with \spherex's consistently being higher due to the much larger effective aperture of its measurements compared to the $3\arcsec\times 3 \arcsec$ FOV of the JWST NIRSPEC IFU (approximately 1/4 the area of one \spherex\ pixel).

\begin{figure*}
   \centering
   \includegraphics[width=0.7\textwidth]{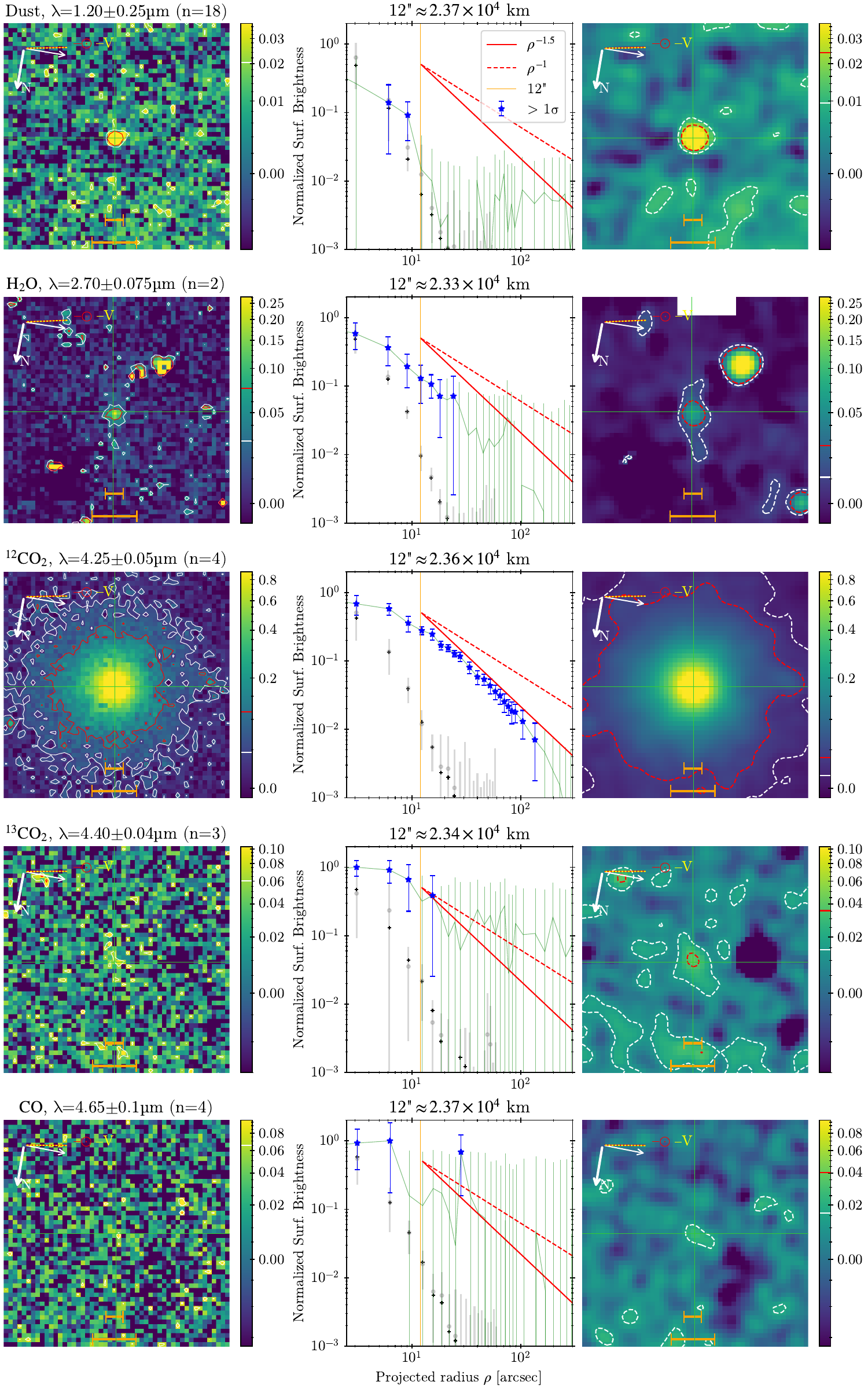}
   \caption{
   Median combined images ($\approx 5\arcmin \times 5\arcmin$) and radial profiles. \textbf{Left column}: The stacked images, after simulation and background subtractions, in MJy/sr unit (see scale bars). The orange bars at the bottom indicate aperture diameter ($24\arcsec$) and $1 \arcmin$. \textbf{Middle column}: The radial profile of \atlas\ in the stacked image (green; blue if the signal is above the 1-$\sigma$ scatter), mean (gray) and median (black) of 30 brightest field stars within $\lesssim 10\arcmin$, respectively. Each profile is normalized to the central pixel. The red solid and dashed lines are for $\rho^{-1.5}$ and $\rho^{-1}$ profiles, respectively, for the radius $\rho$. The nominal aperture radius (orange vertical lines) and corresponding length at \atlas\ (title) are shown.
   \textbf{Right column}: Same as the left column but convolved with a standard deviation of 1.5-pixel gaussian filter for visualization. In the left and right columns, celestial North (white thick arrow), East (white thin arrow), anti-solar (red solid) and anti-velocity (yellow dashed) vectors are shown in the top left corner. The white and red contours (also shown in the color bar to the right of each panel) are 2- and 5-$\sigma$ above the background, respectively.
   }
   \label{fig:stack}
\end{figure*}

Because the $\co_2$ coma is definitely extended beyond $1\arcmin$ (Fig. \ref{fig:stack}), the nominal aperture photometry will give only a lower limit of the flux. Integrating the total flux using the radial profile is discussed below.

\subsection{Gas Production Rates}\label{ss:Qgas}
We adopt a production rate $Q_\mathrm{gas}$ via
\begin{equation}
        Q_\mathrm{gas} = 2 \pi v_\mathrm{gas} \rho N_\mathrm{gas}
\end{equation}
where $N_\mathrm{gas}$ is the average column density of the gas within a circular aperture
of projected radius $\rho$ on the sky, centered on the nucleus, and where $v_\mathrm{gas}$
is the gas emission velocity. We use $v_\mathrm{gas} = (0.85\unit{km/s})/\sqrt{r_\mathrm{hel}\text{ in au}} = 0.48 \unit{km/s}$ \citep{2012Icar..218..144C}
and $r = 11,600$ km here. The average column density is given by
\citep[cf. e.g.][]{1989AdSpR...9c.163K}
\begin{equation}
    N_\mathrm{gas} = \frac{4 r_\mathrm{obs}^2}{\rho^2} \frac{F_\nu \Delta \lambda}{g h \lambda} \left ( \frac{r_\mathrm{hel}}{1\au} \right ) ^2
\end{equation}
where $F_\nu$ is the average flux density (i.e. power per area per frequency bandwidth) in the band from the gas; $r_\mathrm{hel} = 3.2\au$ and $r_\mathrm{obs} = 2.6\au$ are the heliocentric and observer-centric distances to \atlas, respectively; $g$ is the fluorescence efficiency at 1 au for the emission band in question, and we use $g = 2.9\times 10^{-4}$, $2.6\times 10^{-3}$, and $2.6\times 10^{-4}$ photons per second per molecule for $\water$ $2.85\um$, $\co_2$ $4.25\um$, and $\co$ $4.7\um$ emission bands, respectively \citep{1997Sci...275.1904C}; $h$ is
Planck's constant; $\Delta \lambda$ is the bandwidth for the narrow-band assumption, and we use $0.08\um$ for $\water$ and $0.037\um$ for all others; and $\lambda$ is the wavelength of the band.

Given the \spherex\ $\water$, $^{12}\co_2$, $^{13}\co_2$, and CO gas emission fluxes of 58, 8610, 70, and 59 mJy measured in 2 pixel radius effective apertures and corrected for \atlas's relative position to $r_\mathrm{hel} = 1\au$, $r_\mathrm{obs} = 1\au$ (Fig.~\ref{fig:mjy+refl}), we derive gas production rates of
$Q_\mathrm{\water} = 3.2 \times 10^{26} \unit{molec/sec}  \pm 20 \pct$,
$Q_\mathrm{^{12}\co_2} = 1.6 \times 10^{27}  \unit{molec/sec} \pm 10\% $,
$Q_\mathrm{^{13}\co_2} = 1.3\times10^{25} \unit{molec/sec} \pm 25\% $, and
$Q_\mathrm{\co} = 1.0\times10^{26} \unit{molec/sec} \pm 25\pct$.
The rates of $\co_2$ and $\water$ emission are consistent with the activity of inbound thermally processed short-period Solar System comets at $3.2\au$ \citep{2022PSJ.....3..247H}, where water gas emission is ``not fully on yet'', and the ratio of $Q_\mathrm{^{13}\co_2} / Q_\mathrm{^{12}\co_2} \sim 1/100$ is consistent with the ISM value 0f 70,  although the $\co$ production rate at $< 0.1 Q_{\co_2}$ is very low (see Discussion section). These updated Q$_{gas}$ values are also consistent with the preliminary values reported for the first 70 visits by \cite{2025RNAAS...9..242L}, and by \cite{2025ApJ...991L..43C} from their JWST observations of 06-Aug-2025 UT taken roughly halfway through the \spherex\ observing run.

\subsection{Images}
We extracted cutout images at the specified wavelengths, with the few pixels flagged as cold, hot, non-functional, or otherwise erroneous masked. The masked images were then median-combined, and the background removed by subtracting the smoothed 10 by 50 pixels rectangular regions, centered at 75 pixels away to the left and right of the center of \atlas. The resulting stacked, background removed \spherex\ near-infrared images of \atlas\ are shown in Fig. \ref{fig:stack}.

 Images at different wavelength ranges probe the distribution of dust and important spectral features by molecules (Sect. \ref{ss:spec}). The large extension seen in $^{12}\co_2$ versus $\water$ is not surprising given the orders of magnitude higher measured flux emission from $\co_2$ gas, the object's location well within the $\sim 12\au$ $\co_2$ ice line of the solar system, and the long lifetime ($\sim 6$ days) vs photolysis and SWCX compared to water ($\sim 1$ day). However, the falloff of the $\co_2$ coma with respect to projected radial distance $\rho$ is definitely \textcolor{black}{steeper} than $1/\rho$, implying that there is either acceleration of the outflow occurring or there is destruction of the outflowing $\co_2$ going on.

The lack of a bright water gas coma is puzzling as \atlas\ was not far outside the Solar system's ``water ice line'' at $r_\mathrm{hel} = 2.5\au$ during the observations. We are led to conclude that \atlas\ was emitting large chunks of mixed $\co_2$+$\water$ ice into its coma at the time of observation, and that evaporative cooling of $\co_2$ was pinning the chunks' temperature at $\sim120\unit{K}$ and greatly suppressing the $\water$ ice's vapor pressure \citep{2021Icar..35614072L}. A corollary of this argument is that once \atlas\ comes well within the water ice line of the solar system at $2.5 \au$, its outgassing behavior will markedly change and increase as majority matrix water ice itself starts evaporating rapidly. This will destroy many of the large chunks of ice in the coma, changing it over to a small particle dominated cloud with a more anti-solar pointing tail, while driving the sublimation front inward towards the nucleus.

\subsection{Radial profiles}

Azimuthally averaged radial profiles for each stacked image were constructed by taking the modal value of 1-pixel-thick circular annuli after 3-$\sigma$ clipping \citep{1996A&AS..117..393B}. Resulting observed radial profiles representative of dust, H$_2$O, $^{12}$CO$_2$, $^{13}$CO$_2$, and CO, are shown in (Fig. \ref{fig:stack}). The profiles were then compared to those found for nearby stars in the \spherex\ all-sky survey, for determination if \atlas\ appeared extended versus a point source.  If so, we further compared them to the canonical 1/$\rho$ behavior expected for a nucleus emitting material at constant velocity into a $4\pi$ sphere
\citep{1990pcc..conf...69A}.

Only at wavelengths characteristic of emission from $\water$ and  $^{12}\co_2$ do we see significant extension above stellar; the profile for $\water$ is noisy enough that we \textcolor{black}{cannot} say much more about its power-law slope. By contrast, the $^{12}\co_2$ profile slope is very well defined, and clearly falls off more quickly than a $\rho^{-1}$ coma, more like a $\rho^{-3/2}$ coma. \textcolor{black}{Assuming that the CO$_2$ coma is optically thin, (valid for the observed Q$_{CO2}$ = 1 x 10$^{27}$ emission rate into r >6")} this is indicative of either significant photolytic + charge exchange destruction of the $\co_2$, or significant acceleration of the $\co_2$. Given that the $\co_2$ coma extends out to $\sim3\arcmin = 384,000 \km$ at $3.2 \au$, and $\co_2$ gas molecules with a photolytic lifetime of $\sim 6 \unit{days} = 526,000 \unit{s}$ \textcolor{black}{must move at an average speed of $0.7 \unit{km/sec}$, it is hard to see how many of them could be getting destroyed close to the nucleus. Instead they are likely accelerating due to significant pressure in the gas releasing coma, an explanation supported by noting that the \citep{2012Icar..218..144C} gas velocity relation has the molecules emitted thermally at 0.48 km/sec (Section 4.3), and that the $\co_2$ coma gas profiles of Fig. 3 appear quite flat \textcolor{black}{for the first $\sim 15 \arcsec$ ($32,000 \km$) before bending } past a sublimation front to follow a $\rho^{-3/2}$ trend at larger distances.}

\subsection{Nucleus Size\label{sec_nucleus}}
Since the scattered light signal from \atlas\ is unresolved spatially versus a point source by \spherex, we analyze its importance by taking limiting cases. Firstly, if we assume all the flux were to originate as sunlight scattered off of a nuclear surface of geometric albedo $p_\mathrm{V} = 0.04$ at the phase angles $\alpha = \range{16.3}{17.2}\degr$ of \atlas\ during the observation, we find an effective spherical radius of the object of $\sim 23 \km$ \citep{2025RNAAS...9..242L}. However, we know this assumption is patently false, as Hubble Space Telescope observations of the object have placed a firm upper limit, after fitting and tracing the coma flux back to the central PSF pixels, of no more than an $2.5 \km$ radius, $p_\mathrm{V} =0.04$ point source residual. Further, no significant lightcurve has ever been measured for \atlas, and a large asymmetric rotating nucleus, as all comets observed to date have shown, will produce significant lightcurve variations. And finally, \textcolor{black}{the giant $\co_2$ coma mapped by \spherex\ shows a flat-topped structure within $15\arcsec$ (2.5 pixels or 32,000 km) of the nucleus}. 

The solution to all these issues is the same: the presence of a small, rotating, dark nucleus surrounded by a stable, approximately 100 times brighter coma. (Cometary comae are typically static in brightness over hours to days unless the nucleus produces a huge outburst of emission or has a bright focused region of jet-like activity on its surface). Solar system comets where the nucleus is greatly outshown by its coma are rare but not unheard of; they are the handful of small, ``hyperactive'' comets like 41P/Wirtanen and 103P/Hartley 2 (\hartley, hereafter) with so much icy dust in their comae that the coma solid material has more surface area than the nucleus and out-produces the nucleus in gas emission. These comets thus require high amounts of large, long lived icy particle emission, and further appear to be $\co_2$ gas dominated  (\citep{2009PASP..121..968L}; see the Discussion Section for further arguments for the connection between \atlas\ and \hartley). This was just the kind of behavior \atlas\ was exhibiting during the early August 2025 \spherex\ observations reported here.

\section{Discussion} \label{s:disc}

All of the observations reported here (the giant $\co_2$ coma, the lack of any observed lightcurve variability, the prevalence of large coma grains, and the 100 times greater coma scattered light flux than a $2.5 \km$ radius, 0.04 geometric albedo spherical object would produce) are consistent with an object spectrally dominated by bright, $\co_2$- and water ice-rich coma material. The low amount of $\water$ gas production observed by \spherex\ in August 2025 can be explained by the object's having only recently arrived from the cold depths of interstellar space and its remoteness from the Sun (outside the water ice line between $3.4$ and $3.1 \au$) during the observational period \citep{2021Icar..35614072L, 2022PSJ.....3..112L}. The lack of strong $\co$ emission is slightly surprising, as the $\co$-ice line in the solar system lies at $\sim 40 \au$ from the Sun, until one realizes it is an indication of hypervolatile depletion.

\begin{figure*}
\gridline{
    \fig{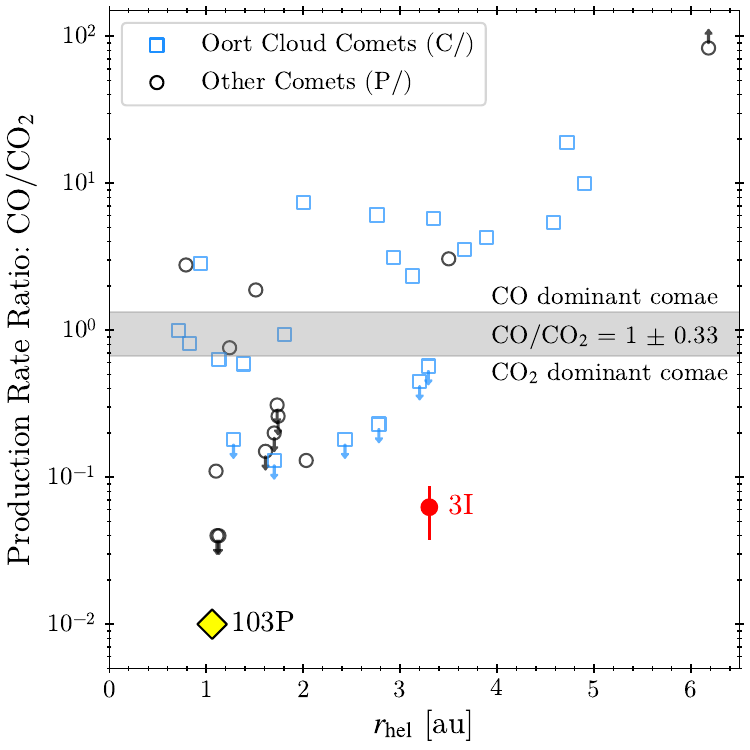}{0.45\textwidth}{(a)}
    \fig{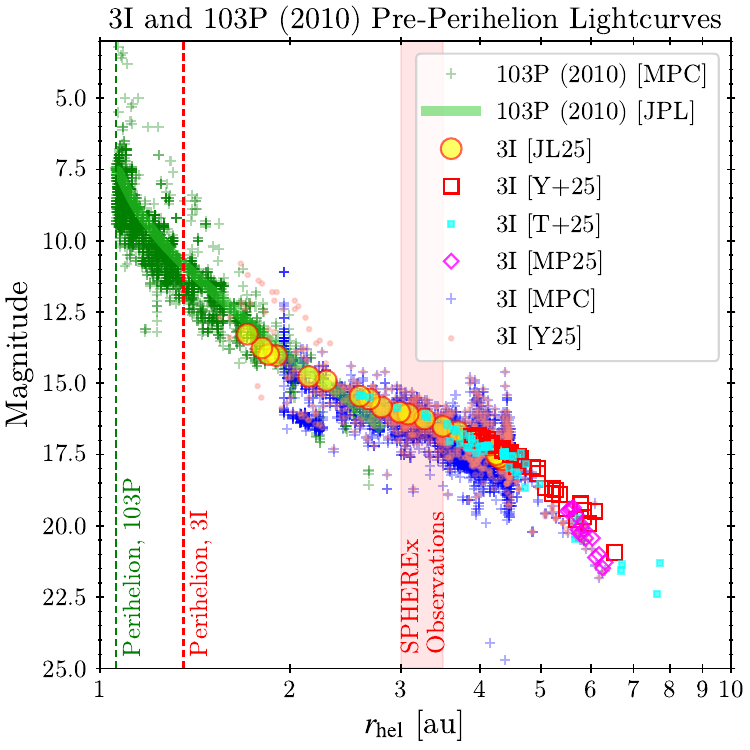}{0.45\textwidth}{(b)}
}
\gridline{
    \fig{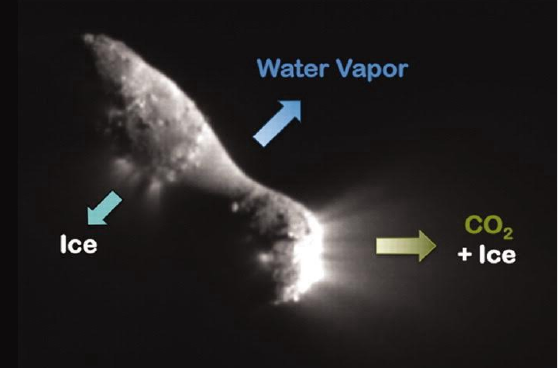}{0.53\textwidth}{(c)}
    \fig{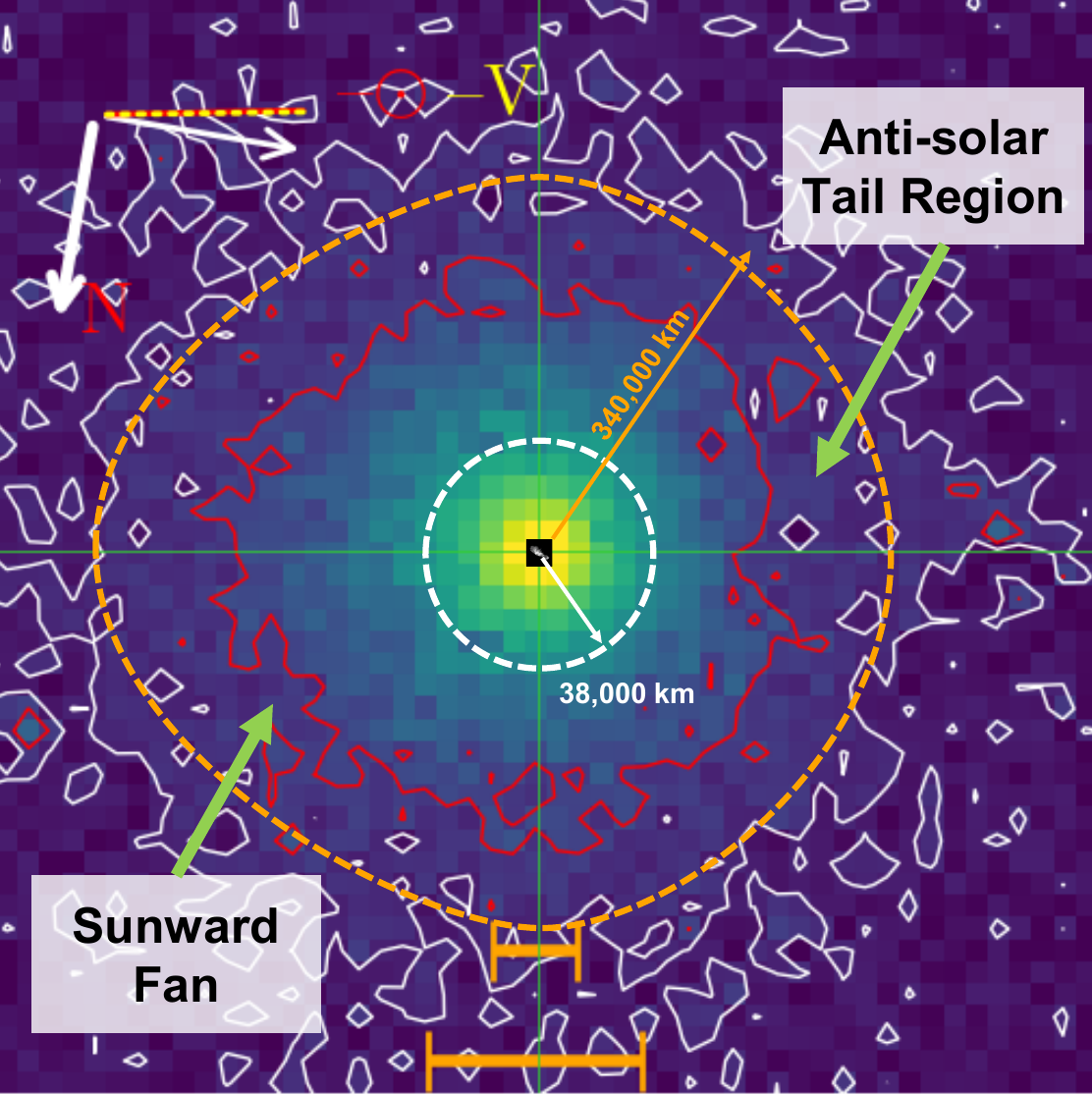}{0.35\textwidth}{(d)}
}
\caption{\textbf{(a)} $\co/\co_2$ gas production rate ratios of solar system comets \citep{2022PSJ.....3..247H}. The $\co$ or $\co_2$ dominance criteria of $1 \pm 0.33$ is indicated as horizontal shadow \citep{2022PSJ.....3..247H}. The ratio of \atlas\ 
makes it extremely $\co$ poor and equivalent to the most thermally processed comet in the 2022 survey, \hartley.
\textbf{(b)} Pre-perihelion optical lightcurves observed for comet \hartley\ during the 2010--2011 apparition and \atlas  in May - Oct 2025.  Published data from \citep{2025ApJ...994L...3J, 2025arXiv250905562T, 2025ApJ...993L..31Y} have been shifted to match the absolute magnitude of Minor Planet Center (MPC) Database\footnote{
  \url{https://www.minorplanetcenter.net/db_search/show_object?object_id=3I} for \atlas\ and \url{https://www.minorplanetcenter.net/db_search/show_object?object_id=103P} for \hartley.
}, the T-mag from JPL Horizons ephemerides service\footnote{\url{https://ssd.jpl.nasa.gov/horizons/app.html}}, and the data kindly shared by S. Yoshida (2025\footnote{\url{http://www.aerith.net/comet/catalog/0003I/2025N1.html}}, priv. comm.). There is very good agreement between the two inbound lightcurve trends. An inherent spread of $\sim 0.5 \unit{mag}$ in the measurements between nights and between observers can be seen.
\textbf{(c)} Close flyby imaging of the $\sim 1.5 \times 0.6 \times 0.6 \km$ comet \hartley\ nucleus as obtained by the Deep Impact Extended mission \citep{2011Sci...332.1396A}. A marked bilobate nuclear morphology was seen, with unusually large (mm to dm) chunks of $\co_2$ rich ice and gas emanating from the end of the smaller lobe and the middle of the larger lobe. The surrounding coma, filled with large long lived icy grains, was found to be at least 5 times brighter and more actively emitting gas than the nucleus. While the nucleus of \atlas\ has never been resolved, given its similar behavior, it could possibly share similar morphological characteristics.
\textbf{(d)} Schematic of the \atlas\ object derived from \spherex\ $\co_2$ photometric mapping. The $< 2.5 \unit{km}$ radius nucleus \citep{2025ApJ...990L...2J} was not resolved at the \spherex\ pixel scale of $6.15 \arcsec \approx 11,600 \unit{km}$ at $2.6 \au$ distance, but a notional nucleus image has been included at the origin to give the reader its location and a sense of scale. Two distances have been marked off: The $\sim 38,000 \unit{km}$ (white dashed circle) inside of which we think coma ice is actively sourcing $\co_2$ by sublimation; and $\sim 340,000 \unit{km}$ (orange dashed oval), the minimum radial distance to which we detect $\co_2$ gas at 2-$\sigma$ level.
}
\label{fig:hartley}
\end{figure*}

Given the current turbulent mix of information currently posted on \atlas\, it is important to state that there are known, studied natural solar system comets producing analogous behavior. A strong predominance of $\co_2$ gas emission coupled with $\co$ hypervolatile depletion has been seen in solar system hyperactive comets like \hartley\ \citep[][$Q_{\co_2}/Q_{\water} \sim 20 \pct$; $Q_{\co}/Q_{\water} \sim 0.15 \pct$]{2012ApJ...758...29A} and 46P/Wirtanen \citep[][$Q_{\co_2}/Q_{\water} \sim 15\pct$; $Q_{\co}/Q_{\water} < 0.5 \pct$]{2021PSJ.....2...21M}. In fact, \hartley\ is the comet with the lowest known ratio of $\co$ to $\co_2$ \citep[][see Fig. \ref{fig:hartley}a]{2022PSJ.....3..247H}. Hyperactive comets are defined as comets with gas production rates many times what sublimation from their nucleus surfaces can/should be able to support; their gas production is instead produced primarily from large, long-lived icy grains emitted into the surrounding coma by the nucleus.

The connection between \atlas\ and \hartley\ is bolstered by reports of \atlas's abundant C2 and CN emission, putting it akin to the spectral Class A comets of Fink 2009's spectral survey, which also contains hyperactive comets \hartley\ and 46P/Wirtanen. Further, comparison of the inbound visible lightcurve vs heliocentric distance for the \hartley\ 2010 apparition to the current inbound lightcurve vs distance for \atlas\ (Fig. \ref{fig:hartley}b) shows an excellent match (Figure 4a).

\hartley\ was especially well studied during the close ($\sim 1000 \km$ minimum distance) flyby of the NASA Deep Impact Extended mission \citep[][in November 2010]{2011Sci...332.1396A}, which imaged large chunks of $\co_2$ rich material being ejected ``in a reverse snow'' from one of the comet's lobes and measured a coma $\sim 5\times$ brighter than the nucleus in total scattered light flux (Figure 4c). The large chunks of $\co_2$-rich material are explained as primordial remnants of solid $\co_2$ ice which have lost much of their more hypervolatile ice phases (e.g., $\co$, $\mathrm{N_2}$, and $\mathrm{CH_4}$). The imaging results from the Deep Impact extended mission suggest that the regions of its nucleus emitting $\co_2$-rich ice are separated from regions producing water gas, and the source ices may be physically separated.

\subsection{Physical Behavior Hypotheses}
\textcolor{black}{There are several hypotheses that have been proposed to explain \atlas's physical behavior. For example, it has been proposed that \atlas\ originated in a very $\co$-poor but $\co_2$-rich exosystem (i.e., the ``nature hypothesis'' to the following ``evolutionary nurture proposals''). Assuming that the $\co:\co_2$ ratio is directly reflective of the natal system's C:O abundance ratio \citep[c.f.,][]{2020NatAs...4..867B}, we find this explanation problematic, as this would imply a system with a host star C/O ratio $< 1/10$ the Sun's, and a dearth of reduced aliphatic and aromatic species as well. [C/Fe] and [C/H] $< -1.0$ for a nearby Milky Way star is highly improbable \citep{2017ApJ...848...34H, 2017LPICo2042.4061L, 2025AJ....170...96M}, and the aliphatic and PAH absorption features for \atlas\ in the JWST NIRSPEC spectrum of \cite{2025ApJ...991L..43C} look to be of reasonably normal strength.}

\textcolor{black}{Another hypothesis from the August 2025 observations suggests that \atlas\ is in the same state of advanced thermal processing, mall loss, and age as analogously behaving \hartley. Because of \hartley's very small size but moderate to high gas production rate, coupled with its lack of hypervolatile $\co$ and presence of large $\co_2$ -rich grains, it has been argued that it has been highly thermally processed by the Sun over hundreds to thousands of orbits, and is near the end of its life \citep{2009PASP..121..968L}. To get to this state, it has lost much of its original mass, so that its current surface used to be its core material, buried under kilometer's worth of cometary material at its creation.}

\textcolor{black}{Another possibility is that we are observing a comet-like body from an exosystem with near-solar composition, but also a body with unusual alteration of its surface layers by Gyrs of continual galactic cosmic ray and stellar XUV irradiation kept at $T_\mathrm{galactic\ ISR} = \range{15}{30}\unit{K}$ \citep{2014A&A...564A.111ZB}. This would produce an $\co:\co_2$ number ratio of $\approx 0.1$ if the surface material started out as $\water:\co_2$ ice only, without any starting $\co$ \citep{2025arXiv251026308M}. However, if $\co$ was initially present in the newly ejected comet’s surface at typical primordial-like abundance ratios of $\co:\co_2 = \range{0.5}{10}$, the irradiated result should have a final $\co:\co_2$ ratio $\sim 1$  (Moore 1991). The observed \spherex\ $\co:\co_2$ ratio of $0.06 \pm 0.03$ is consistent with a predicted value of $0.1$ for a starting pure $\water:\co_2$ ice case, and the large emitted dust grains are reminiscent of the mantle grains emitted from surface processed short-period solar system comets \citep{2002EM&P...90..497L} implying that \atlas\ was thermally processed enough in its home system to severely deplete hypervolatiles in its near-surface layers before ejection and irradiation. }

\textcolor{black}{The way to distinguish between the [Complete thermal $\co$ depletion of an object] versus [Surface $\co$ thermal depletion followed by meters worth of galactic radiation treatment] hypotheses is to search for a marked change in the relative $\co$ emission behavior of \atlas\, both in increased abundance and direct nuclear emission morphology, as it passes close to the Sun on 30 Oct 2025 and its interior regions, beneath its $\sim 10 \unit{m}$ surface irradiation processed zone \citep{2025arXiv251026308M}, become heated.}

\subsection{ISO Evolution and Home System Implications} 
A strongly thermally processed state for \atlas\ has important implications. By analogy to \hartley's behavior, it tells us that \atlas\ has undergone 100s to 1,000s of close passages by its natal host/birth star (we reject the possibility that \atlas\ has undergone hundreds to thousands of close passages of random galactic stars on its passage to the solar system; this rate of close passage for a single object, even one as old as the galaxy, would imply a huge rate of interstellar object passages \& detections in our solar system, which is not seen). 

These close thermally altering passages could have been accomplished either as a primordial body formed in-place near its birth system's $\co_2$ ice line (to ensure minimal CO but extensive labile  $\co_2$ ice incorporation) and quickly ejected soon after PPD disk clearing, or, as seen currently in our own mature system, as an ice-rich KBO that became scattered into its inner system for a few hundred near-star orbits before ejection.

\textcolor{black}{Finally we end with a piece of scholarly speculation. Naive models of ISO creation would suggest the maximal number of ISOs created by a system should be when it hosts the most free-flying small planetesimals that are dynamically hot. I.e., when the system is very young, within the first few Myrs of the host stars formation. But this is also the era when a system's protoplanetary disk is optically thick near its midplane and planetesimals in the densest part of the disk remain cold and thermally unprocessed. PPDs are seen to clear within $\range{5}{10} \unit{Myrs}$ of host star formation though \citep[and references therein]{2005ApJ...620.1010R, 2025MNRAS.541.2246B},
and the first heavy thermal processing of KBOs estimated to occur in $\range{1}{40} \unit{Myrs}$
\citep{2021MNRAS.505.5654D, 2021Icar..35613998S, 2021Icar..35614072L, 2022PSJ.....3..251L, 2023MNRAS.522.2081P}. }

Planetary migration and Oort Cloud formation involving small body ejection processes can take 100's of Myrs to complete \citep{2013Icar..225...40B, 2019ApJ...884...69G}.   We can thus plausibly argue that PPDs can clear and small body thermal evolutionary processes can occur more rapidly than dynamical processes can cool down and mature in a system, and we should see a mix of volatile rich, hypervolatile depleted, and totally devolatilized ejected objects as our ISOs. This is precisely what we have seen so far in highly-devolatilized 1I/Oumuamua,  hypervolatile-rich 2I/Borisov, and well-thermally processed \atlas.

\section{Conclusions and Future Work} \label{sec:conc}

In conclusion, the \spherex\ spacecraft observed ISO \atlas\ from 01 to 15-Aug-2025 UT using 102 band, $R = \range{40}{130}$ spectrophotometry. Photometric imagery, spectroscopy, and light curves of the ISO were obtained. From these, robust detections of slightly extended water gas emission at $\range{2.7}{2.8} \um$ and highly extended  $^{12}\co_2$ gas at $\range{4.23}{4.27} \um$ were found. More tentative detections of $^{13}\co_2$ and $\co$ gas were seen. A slightly extended coma of $\water$ was detected, and a huge surrounding atmosphere of $\co_2$ extending out to at least 3' was discovered. No jets were detected in the coma imagery, but a slight sunward pointing asymmetry was seen in the $\co_2$ coma. Derived gas production rates for the 4 species were 
$Q_\water       = 3.2 \times 10^{26} \pm 20 \pct$, $Q_{^{12}\co_2} = 1.6 \times 10^{27} \pm 10 \pct$, $Q_{^{13}\co_2} = 1.3 \times 10^{25} \pm 25 \pct$, and 
$Q_\co          = 1.0 \times 10^{26} \pm 25 \pct$.

Registration and co-adding of dozens of scattered light continuum images at $\range{1.0}{1.5} \um$ produced a high S/N ratio image consistent with a stellar point source of no extension. The same images were used to create a lightcurve, for which variability $\lesssim 15 \pct$ was seen from beginning to end of the observing period. The absolute brightness of \atlas\ at $\range{1.0}{1.5}\um$ is consistent with a nucleus of effective radius $R_\mathrm{nuc} \lesssim  2.5 \unit{km}$, as reported by HST \citep{2025ApJ...990L...2J} surrounded by a 100 times brighter coma. The \spherex\ spectral continuum measured over 2 weeks is highly consistent with the continuum reported by JWST during its single visit to the object on 06 Aug 2025 \citep{2025ApJ...991L..43C}. The $\range{1.5}{4.0} \um$ continuum structure shows a strong feature commensurate with water ice + refractory organics absorption. The overall shape of this reflectance continuum is very similar to that reported for the cliff-type KBOs by \citep{2025NatAs...9..230P}

The August \spherex\  observations provided the first detection of strong $\co_2$ emission from the comet into a giant extended surrounding coma, proving that $\co_2$ sublimation from large icy coma grains is the main driver for the observed outgassing activity and explaining its extended morphology as the product of $\co_2$ gas sublimation that was undetectable in the first reported UVIS characterizations of the object. The moderate observed water gas production rate can be explained as an object just coming in from the cold depths of interstellar space that was observed at $r_\mathrm{hel} = 3.2 \au$ from the Sun outside the water ice line.

In many ways the observed activity behavior of \atlas, including its preponderance of $\co_2$ emission, lack of $\co$ output, small size, and predominance of large chunks of water-ice rich icy material in a flux-dominant coma is reminiscent of the behavior of short period comet \hartley, target of the NASA Deep Impact extended mission in 2010 and a ``hyperactive comet''  \citep{2011Sci...332.1396A, 2009PASP..121..968L}. Such an object could have undergone hundreds to thousands of close perihelion passages by its natal birth star inside the system's water ice line, as \hartley\ has by our Sun. Whether this was due to \atlas\ forming within a few au of its birth star or being scattered into this region after formation but before ejection is not clear. 

Compared to the naïve expectation that solar systems shed the majority of their comets in their first few Myrs of existence when their small icy planetesimals are most abundant, most hypervolatile ice rich, and most dynamically hot, so that all ISOs should be like 2I/Borisov, this suggests that ISOs are thermally processed quicker than they dynamically relax. This hypothesis needs to be tested by population studies of many more ISOs than the 3 currently known, and hopefully many more will be studied by the \spherex\ mission's all-sky survey.

\begin{acknowledgments}

This paper relies on data obtained by the \spherex\ Observatory (operated by Caltech and JPL on behalf of the National Aeronautics Space Administration under contract 80GSFC18C0011) that is hosted by IPAC as part of the IRSA archive (10.26131/IRSA652 \spherex\ Quick Release Spectral Images - QR2). We are indebted to the \spherex\ project for obtaining additional pointings at \atlas, approximately 100 percent more, than their nominal survey would have obtained less than 3 months after the start of the survey by testing and implementing a new commanding protocol. We have also used optical lightcurve trending data kindly supplied by S. Yoshida. The research was carried out at the Jet Propulsion Laboratory, California Institute of Technology, the Johns Hopkins Applied Physics Laboratory, and Arizona State University under a contract with the National Aeronautics and Space Administration 
(80NM0018D0004). The authors also acknowledge the Texas Advanced Computing Center (TACC) \footnote{\url{http://www.tacc.utexas.edu}} at The University of Texas at Austin for providing computational resources that have contributed to the research results reported within this paper.

\end{acknowledgments}

\begin{contribution}

All listed authors contributed importantly to the creation and production of this manuscript.
\end{contribution}

\facilities{SPHEREx, IRTF, SOLO}

\software{
SPHEREx-Sky-Simulator \citep{2025ApJS..281...10C},
spiceypy \citep{2020JOSS....5.2050A}, SPICE \citep{2018P&SS..150....9A, 1996P&SS...44...65A},
kete \citep{2025arXiv250904666D},
skyloc\footnote{\url{https://github.com/ysBach/skyloc}},
SPHEREx-SSO,
astropy\citep{2013A&A...558A..33A,2018AJ....156..123A,2022ApJ...935..167A}, ccdproc \citep{2025zndo....593516C},
photutils \citep{2022zndo....596036B},
}

\appendix

\section{Appendix information}

\bibliography{3i_atlas,spherex_bibs}{}
\bibliographystyle{aasjournal}

\clearpage
\startlongtable

\begin{deluxetable}{cccccccccccc}
\tablewidth{0pt}
\tablecaption{Observation log\label{tab:obs}}
\tablehead{
\colhead{ObsID}
& \colhead{$\lambda$}
& \colhead{Mid JD}
& \colhead{Flag}
& \colhead{$F_\lambda$}
& \colhead{$dF_\lambda$}
& \colhead{$r_\mathrm{hel}$}
& \colhead{$r_\mathrm{obs}$}
& \colhead{$\alpha$}
& \colhead{$F_{\lambda, 1}$}
& \colhead{Refl}
& \colhead{$F_\mathrm{field}$} \\
\colhead{} & \colhead{$\mathrm{\mu m}$} & \colhead{UTC} &\colhead{} & \colhead{$\mathrm{mJy}$} & \colhead{$\mathrm{mJy}$} & \colhead{$\mathrm{au}$} & \colhead{$\mathrm{au}$} & \colhead{$\mathrm{{}^{\circ}}$} & \colhead{$\mathrm{mJy}$} & \colhead{} & \colhead{}
}
\startdata
2025W33\_1B\_0134\_1 & 0.7543 & 2460899.2257 & b & 1.5806 & 0.1298 & 3.146 & 2.689 & 17.89 & 113.0997 & 0.9117 & 0.66 \\
2025W33\_1B\_0576\_1 & 0.7555 & 2460901.6032 &  & 1.6693 & 0.1128 & 3.070 & 2.669 & 18.70 & 112.0514 & 0.9028 & 0.00 \\
2025W33\_1B\_0134\_2 & 0.7720 & 2460899.2272 & b & 1.6806 & 0.1290 & 3.146 & 2.689 & 17.89 & 120.2481 & 0.9617 & 0.68 \\
2025W33\_1B\_0576\_2 & 0.7732 & 2460901.6047 &  & 1.4010 & 0.1102 & 3.070 & 2.669 & 18.70 & 94.0379 & 0.7516 & 0.05 \\
2025W33\_1B\_0134\_3 & 0.7904 & 2460899.2286 & b & 1.4530 & 0.1281 & 3.146 & 2.689 & 17.89 & 103.9565 & 0.8258 & 0.72 \\
2025W33\_1B\_0576\_3 & 0.7914 & 2460901.6062 &  & 1.7357 & 0.1133 & 3.070 & 2.669 & 18.70 & 116.4995 & 0.9247 & 0.05 \\
2025W33\_1B\_0183\_1 & 0.8086 & 2460899.4981 &  & 1.5183 & 0.1098 & 3.137 & 2.686 & 17.98 & 107.8433 & 0.8554 & 0.03 \\
2025W33\_1B\_0576\_4 & 0.8102 & 2460901.6077 &  & 1.6141 & 0.1113 & 3.070 & 2.669 & 18.70 & 108.3380 & 0.8588 & 0.11 \\
2025W33\_1B\_0183\_2 & 0.8277 & 2460899.4996 &  & 1.5842 & 0.1100 & 3.137 & 2.686 & 17.98 & 112.5217 & 0.8892 & 0.03 \\
2025W33\_1B\_0183\_3 & 0.8469 & 2460899.5011 &  & 1.6796 & 0.1074 & 3.137 & 2.686 & 17.98 & 119.2890 & 0.9370 & 0.16 \\
2025W33\_1B\_0183\_4 & 0.8669 & 2460899.5026 & b & 1.6537 & 0.1115 & 3.137 & 2.686 & 17.99 & 117.4487 & 0.9180 & 0.34 \\
2025W33\_1B\_0447\_1 & 0.8825 & 2460900.9246 &  & 1.6678 & 0.1088 & 3.092 & 2.674 & 18.47 & 114.0067 & 0.8880 & 0.01 \\
2025W33\_2C\_0073\_1 & 0.8880 & 2460902.2831 & b & 3.5835 & 0.1330 & 3.049 & 2.663 & 18.92 & 236.2225 & 1.8377 & 0.33 \\
2025W33\_2C\_0255\_1 & 0.9028 & 2460903.0304 & b & 1.6838 & 0.1201 & 3.025 & 2.657 & 19.16 & 108.8071 & 0.8427 & 0.57 \\
2025W33\_1B\_0447\_2 & 0.9034 & 2460900.9261 &  & 1.7800 & 0.1116 & 3.092 & 2.674 & 18.47 & 121.6712 & 0.9423 & 0.01 \\
2025W33\_2C\_0220\_1 & 0.9105 & 2460902.8942 & a & 1.6894 & 0.1147 & 3.029 & 2.658 & 19.12 & 109.5633 & 0.8453 & 0.31 \\
2025W32\_2D\_0224\_1 & 0.9184 & 2460896.2371 & b & 1.1585 & 0.1155 & 3.242 & 2.718 & 16.81 & 89.9130 & 0.6914 & 0.34 \\
2025W32\_2D\_0224\_2 & 0.9402 & 2460896.2386 & b & 1.1165 & 0.1183 & 3.242 & 2.718 & 16.81 & 86.6476 & 0.6660 & 0.37 \\
2025W32\_2D\_0343\_1 & 0.9621 & 2460896.8510 &  & 1.4387 & 0.1157 & 3.222 & 2.711 & 17.04 & 109.7945 & 0.8448 & 0.03 \\
2025W32\_2D\_0343\_2 & 0.9851 & 2460896.8525 & a & 1.4547 & 0.1163 & 3.222 & 2.711 & 17.04 & 111.0136 & 0.8573 & 0.03 \\
2025W32\_2D\_0343\_3 & 1.0087 & 2460896.8540 &  & 1.5769 & 0.1212 & 3.222 & 2.711 & 17.04 & 120.3376 & 0.9347 & 0.01 \\
2025W33\_1B\_0642\_1 & 1.0100 & 2460901.9458 & b & 2.0724 & 0.1201 & 3.059 & 2.666 & 18.81 & 137.8480 & 1.0715 & 0.42 \\
2025W32\_2D\_0343\_4 & 1.0326 & 2460896.8555 &  & 1.5691 & 0.1165 & 3.222 & 2.711 & 17.04 & 119.7352 & 0.9367 & 0.01 \\
2025W33\_1B\_0642\_2 & 1.0341 & 2460901.9473 & b & 1.9734 & 0.1268 & 3.059 & 2.666 & 18.81 & 131.2523 & 1.0273 & 0.57 \\
2025W32\_2D\_0355\_1 & 1.0578 & 2460896.9169 &  & 1.6547 & 0.1500 & 3.220 & 2.711 & 17.06 & 126.0554 & 0.9919 & -0.00 \\
2025W33\_1B\_0642\_3 & 1.0587 & 2460901.9488 & b & 2.0959 & 0.2123 & 3.059 & 2.666 & 18.81 & 139.3993 & 1.0975 & 0.63 \\
2025W32\_2D\_0355\_2 & 1.0830 & 2460896.9184 & b & 1.6904 & 0.2139 & 3.220 & 2.711 & 17.06 & 128.7693 & 1.0183 & -0.04 \\
2025W33\_1B\_0642\_4 & 1.0839 & 2460901.9503 & b & 1.6923 & 0.2425 & 3.059 & 2.666 & 18.81 & 112.5511 & 0.8901 & 0.66 \\
2025W33\_1B\_0010\_3 & 1.0937 & 2460898.5514 & b & 4.7982 & 0.8953 & 3.168 & 2.695 & 17.65 & 349.6425 & 2.7681 & -0.07 \\
2025W32\_2D\_0213\_4 & 1.0992 & 2460896.1734 & b & 3.7113 & 0.3139 & 3.244 & 2.718 & 16.79 & 288.5384 & 2.2846 & 0.86 \\
2025W32\_2D\_0355\_3 & 1.1095 & 2460896.9199 &  & 1.7655 & 0.1303 & 3.220 & 2.711 & 17.06 & 134.4829 & 1.0649 & -0.01 \\
2025W33\_1B\_0010\_4 & 1.1190 & 2460898.5529 & c & 1.6069 & 0.1423 & 3.167 & 2.695 & 17.65 & 117.0914 & 0.9278 & 0.17 \\
2025W33\_1B\_0020\_1 & 1.1448 & 2460898.6147 &  & 1.7179 & 0.1216 & 3.166 & 2.694 & 17.67 & 124.9685 & 0.9902 & 0.06 \\
2025W33\_1B\_0020\_2 & 1.1722 & 2460898.6162 &  & 1.7996 & 0.1239 & 3.165 & 2.694 & 17.67 & 130.9044 & 1.0379 & 0.10 \\
2025W33\_1B\_0020\_3 & 1.2005 & 2460898.6177 & c & 1.7384 & 0.1259 & 3.165 & 2.694 & 17.68 & 126.4498 & 1.0040 & 0.13 \\
2025W33\_1B\_0020\_4 & 1.2291 & 2460898.6192 & c & 1.4737 & 0.1273 & 3.165 & 2.694 & 17.68 & 107.1920 & 0.8529 & 0.20 \\
2025W33\_1B\_0049\_1 & 1.2567 & 2460898.7511 &  & 1.7138 & 0.1163 & 3.161 & 2.693 & 17.72 & 124.2077 & 0.9960 & 0.09 \\
2025W33\_1B\_0049\_2 & 1.2869 & 2460898.7526 &  & 1.8549 & 0.1162 & 3.161 & 2.693 & 17.72 & 134.4296 & 1.0909 & 0.04 \\
2025W33\_1B\_0049\_3 & 1.3170 & 2460898.7541 &  & 1.5957 & 0.1165 & 3.161 & 2.693 & 17.72 & 115.6397 & 0.9529 & 0.01 \\
2025W33\_1B\_0049\_4 & 1.3483 & 2460898.7556 &  & 1.6960 & 0.1143 & 3.161 & 2.693 & 17.72 & 122.9079 & 1.0275 & -0.02 \\
2025W32\_2D\_0189\_1 & 1.3574 & 2460896.0344 & b & 1.4183 & 0.1227 & 3.248 & 2.720 & 16.74 & 110.6876 & 0.9284 & 0.32 \\
2025W31\_2A\_0219\_1 & 1.3691 & 2460889.1740 &  & 1.1899 & 0.1253 & 3.470 & 2.802 & 14.08 & 112.4649 & 0.9467 & 0.22 \\
2025W32\_2D\_0189\_2 & 1.3899 & 2460896.0359 & b & 1.2936 & 0.1253 & 3.248 & 2.720 & 16.74 & 100.9479 & 0.8545 & 0.35 \\
2025W31\_2A\_0219\_2 & 1.4018 & 2460889.1755 &  & 1.2101 & 0.1256 & 3.470 & 2.802 & 14.08 & 114.3738 & 0.9708 & 0.16 \\
2025W32\_2D\_0189\_3 & 1.4229 & 2460896.0374 &  & 1.4522 & 0.1255 & 3.248 & 2.720 & 16.74 & 113.3258 & 0.9657 & 0.23 \\
2025W31\_2A\_0219\_3 & 1.4354 & 2460889.1770 &  & 1.2089 & 0.1339 & 3.470 & 2.802 & 14.08 & 114.2550 & 0.9758 & 0.27 \\
2025W32\_2D\_0189\_4 & 1.4576 & 2460896.0389 &  & 1.4229 & 0.1225 & 3.248 & 2.720 & 16.74 & 111.0323 & 0.9536 & 0.17 \\
2025W33\_1B\_0599\_1 & 1.4811 & 2460901.7420 &  & 1.6676 & 0.1148 & 3.066 & 2.667 & 18.74 & 111.5249 & 0.9619 & 0.06 \\
2025W33\_1B\_0399\_1 & 1.5216 & 2460900.6541 & b & 1.5960 & 0.1209 & 3.100 & 2.676 & 18.38 & 109.8950 & 0.9569 & 0.43 \\
2025W32\_2D\_0201\_1 & 1.5290 & 2460896.1047 &  & 1.3956 & 0.1240 & 3.246 & 2.719 & 16.76 & 108.7057 & 0.9490 & 0.26 \\
2025W33\_1B\_0399\_2 & 1.5583 & 2460900.6556 & b & 1.6922 & 0.1288 & 3.100 & 2.676 & 18.38 & 116.5121 & 1.0267 & 0.35 \\
2025W32\_2D\_0201\_2 & 1.5663 & 2460896.1062 & b & 1.0992 & 0.1233 & 3.246 & 2.719 & 16.77 & 85.6153 & 0.7579 & 0.37 \\
2025W33\_1B\_0399\_3 & 1.5961 & 2460900.6571 & b & 1.7193 & 0.1282 & 3.100 & 2.676 & 18.38 & 118.3695 & 1.0636 & 0.37 \\
2025W32\_2D\_0201\_3 & 1.6041 & 2460896.1077 & b & 1.2537 & 0.1192 & 3.246 & 2.719 & 16.77 & 97.6421 & 0.8809 & 0.33 \\
2025W33\_1B\_0399\_4 & 1.6346 & 2460900.6586 & b & 1.7319 & 0.1178 & 3.100 & 2.676 & 18.38 & 119.2337 & 1.0961 & 0.35 \\
2025W32\_2D\_0201\_4 & 1.6425 & 2460896.1092 & c & 1.1699 & 0.1173 & 3.246 & 2.719 & 16.77 & 91.1135 & 0.8406 & 0.29 \\
2025W32\_2D\_0480\_4 & 1.6513 & 2460897.5324 &  & 1.3873 & 0.1205 & 3.200 & 2.705 & 17.29 & 103.9303 & 0.9644 & -0.00 \\
2025W32\_1A\_0612\_1 & 1.6882 & 2460894.4706 &  & 1.2311 & 0.1104 & 3.299 & 2.736 & 16.15 & 100.3018 & 0.9585 & -0.07 \\
2025W32\_1A\_0612\_2 & 1.7276 & 2460894.4721 &  & 1.2954 & 0.1092 & 3.298 & 2.736 & 16.15 & 105.5386 & 1.0405 & 0.00 \\
2025W32\_1A\_0637\_1 & 1.7675 & 2460894.6067 &  & 1.1888 & 0.1024 & 3.294 & 2.735 & 16.20 & 96.4935 & 0.9830 & 0.20 \\
2025W32\_1A\_0637\_2 & 1.8089 & 2460894.6082 &  & 1.0186 & 0.1030 & 3.294 & 2.735 & 16.20 & 82.6736 & 0.8736 & 0.26 \\
2025W32\_1A\_0637\_3 & 1.8514 & 2460894.6097 &  & 1.0842 & 0.1022 & 3.294 & 2.735 & 16.20 & 87.9947 & 0.9651 & 0.28 \\
2025W32\_1A\_0637\_4 & 1.8946 & 2460894.6112 & b & 0.9764 & 0.1063 & 3.294 & 2.735 & 16.21 & 79.2467 & 0.9003 & 0.34 \\
2025W32\_2D\_0426\_4 & 2.0255 & 2460897.2641 &  & 1.0247 & 0.1041 & 3.209 & 2.707 & 17.19 & 77.3232 & 0.9774 & 0.28 \\
2025W32\_2D\_0017\_1 & 2.0455 & 2460895.0837 & b & 0.6625 & 0.0992 & 3.279 & 2.730 & 16.38 & 53.0690 & 0.6838 & 0.41 \\
2025W31\_2A\_0156\_1 & 2.0559 & 2460888.8341 & b & 0.6445 & 0.1313 & 3.481 & 2.807 & 13.94 & 61.5102 & 0.8001 & 0.75 \\
2025W32\_1A\_0662\_1 & 2.0912 & 2460894.7445 & b & 0.7503 & 0.0914 & 3.290 & 2.733 & 16.26 & 60.6676 & 0.8144 & 0.38 \\
2025W32\_2D\_0017\_2 & 2.0930 & 2460895.0852 & b & 0.7243 & 0.0919 & 3.279 & 2.730 & 16.38 & 58.0158 & 0.7800 & 0.30 \\
2025W31\_2A\_0156\_2 & 2.1037 & 2460888.8356 & b & 0.8240 & 0.1328 & 3.481 & 2.807 & 13.94 & 78.6373 & 1.0654 & 0.52 \\
2025W32\_1A\_0662\_2 & 2.1400 & 2460894.7460 & b & 0.8510 & 0.0910 & 3.290 & 2.733 & 16.26 & 68.8102 & 0.9590 & 0.32 \\
2025W32\_2D\_0017\_3 & 2.1421 & 2460895.0867 & b & 0.5806 & 0.0939 & 3.279 & 2.730 & 16.38 & 46.5038 & 0.6486 & 0.38 \\
2025W31\_2A\_0156\_3 & 2.1531 & 2460888.8371 & b & 0.7823 & 0.1243 & 3.481 & 2.807 & 13.94 & 74.6552 & 1.0468 & 0.36 \\
2025W32\_1A\_0662\_3 & 2.1903 & 2460894.7475 &  & 0.8273 & 0.0932 & 3.290 & 2.733 & 16.26 & 66.8928 & 0.9626 & 0.18 \\
2025W32\_2D\_0017\_4 & 2.1923 & 2460895.0882 & b & 0.5322 & 0.0913 & 3.279 & 2.730 & 16.39 & 42.6302 & 0.6144 & 0.40 \\
2025W31\_2A\_0156\_4 & 2.2031 & 2460888.8386 & c & 0.7369 & 0.1258 & 3.481 & 2.806 & 13.94 & 70.3206 & 1.0215 & 0.27 \\
2025W33\_1B\_0499\_1 & 2.2173 & 2460901.1996 &  & 0.9691 & 0.0979 & 3.083 & 2.672 & 18.56 & 65.7561 & 0.9640 & 0.04 \\
2025W32\_1A\_0662\_4 & 2.2408 & 2460894.7490 &  & 0.8229 & 0.0939 & 3.290 & 2.733 & 16.26 & 66.5343 & 0.9923 & 0.14 \\
2025W32\_2D\_0091\_2 & 2.2899 & 2460895.4945 &  & 0.8248 & 0.0898 & 3.266 & 2.725 & 16.54 & 65.3319 & 1.0132 & -0.01 \\
2025W32\_2D\_0091\_3 & 2.3421 & 2460895.4960 &  & 0.7588 & 0.0879 & 3.266 & 2.725 & 16.54 & 60.1004 & 0.9709 & 0.01 \\
2025W32\_2D\_0091\_4 & 2.3965 & 2460895.4975 &  & 0.6686 & 0.1167 & 3.265 & 2.725 & 16.54 & 52.9495 & 0.8874 & 0.10 \\
2025W33\_1B\_0576\_1 & 2.4549 & 2460901.6032 &  & 0.7359 & 0.0815 & 3.070 & 2.669 & 18.70 & 49.4018 & 0.8600 & -0.03 \\
2025W33\_1B\_0134\_1 & 2.4555 & 2460899.2257 & b & 0.2977 & 0.0957 & 3.146 & 2.689 & 17.89 & 21.3013 & 0.3708 & 0.90 \\
2025W33\_1B\_0134\_2 & 2.5188 & 2460899.2272 & b & 0.2957 & 0.0841 & 3.146 & 2.689 & 17.89 & 21.1555 & 0.3789 & 0.91 \\
2025W33\_1B\_0576\_2 & 2.5191 & 2460901.6047 &  & 0.5614 & 0.0760 & 3.070 & 2.669 & 18.70 & 37.6839 & 0.6749 & 0.09 \\
2025W33\_1B\_0134\_3 & 2.5870 & 2460899.2286 & b & 0.1190 & 0.0819 & 3.146 & 2.689 & 17.89 & 8.5140 & 0.1611 & 0.96 \\
2025W33\_1B\_0576\_3 & 2.5878 & 2460901.6062 &  & 0.5184 & 0.0765 & 3.070 & 2.669 & 18.70 & 34.7952 & 0.6585 & 0.23 \\
2025W33\_1B\_0183\_1 & 2.6546 & 2460899.4981 & a & 0.6898 & 0.0695 & 3.137 & 2.686 & 17.98 & 48.9947 & 0.9734 & -0.02 \\
2025W33\_1B\_0576\_4 & 2.6581 & 2460901.6077 &  & 0.9513 & 0.0755 & 3.070 & 2.669 & 18.70 & 63.8524 & 1.2686 & 0.15 \\
2025W33\_1B\_0183\_2 & 2.7266 & 2460899.4996 & c & 0.9177 & 0.0704 & 3.137 & 2.686 & 17.98 & 65.1836 & 1.3454 & -0.00 \\
2025W33\_1B\_0183\_3 & 2.7994 & 2460899.5011 & c & 0.3622 & 0.0683 & 3.137 & 2.686 & 17.98 & 25.7283 & 0.5574 & 0.22 \\
2025W33\_1B\_0183\_4 & 2.8755 & 2460899.5026 & b & 0.1377 & 0.0653 & 3.137 & 2.686 & 17.99 & 9.7773 & 0.2224 & 0.68 \\
2025W33\_1B\_0447\_1 & 2.9337 & 2460900.9246 &  & 0.1042 & 0.0659 & 3.092 & 2.674 & 18.47 & 7.1250 & 0.1678 & 0.21 \\
2025W33\_2C\_0073\_1 & 2.9586 & 2460902.2831 & b & -0.0007 & 0.0773 & 3.049 & 2.663 & 18.92 & -0.0490 & -0.0012 & 1.00 \\
2025W33\_2C\_0255\_1 & 3.0131 & 2460903.0304 & b & -0.1549 & 0.0672 & 3.025 & 2.657 & 19.16 & -10.0110 & -0.2476 & 1.18 \\
2025W33\_1B\_0447\_2 & 3.0141 & 2460900.9261 &  & 0.0758 & 0.0646 & 3.092 & 2.674 & 18.47 & 5.1789 & 0.1281 & 0.14 \\
2025W33\_2C\_0220\_1 & 3.0445 & 2460902.8942 & b & -0.1782 & 0.0687 & 3.029 & 2.658 & 19.12 & -11.5603 & -0.2894 & 1.36 \\
2025W32\_2D\_0224\_1 & 3.0721 & 2460896.2371 & b & -0.4158 & 0.0657 & 3.242 & 2.718 & 16.81 & -32.2688 & -0.8275 & 4.59 \\
2025W32\_2D\_0224\_2 & 3.1568 & 2460896.2386 & b & -0.3980 & 0.0673 & 3.242 & 2.718 & 16.81 & -30.8892 & -0.8276 & 3.08 \\
2025W32\_2D\_0343\_1 & 3.2415 & 2460896.8510 &  & 0.1442 & 0.1579 & 3.222 & 2.711 & 17.04 & 11.0012 & 0.3082 & -1.83 \\
2025W32\_2D\_0343\_2 & 3.3290 & 2460896.8525 &  & 0.1357 & 0.1407 & 3.222 & 2.711 & 17.04 & 10.3565 & 0.3044 & -0.19 \\
2025W32\_2D\_0343\_3 & 3.4221 & 2460896.8540 & a & -0.0930 & 0.0888 & 3.222 & 2.711 & 17.04 & -7.0998 & -0.2197 & 0.07 \\
2025W33\_1B\_0642\_1 & 3.4250 & 2460901.9458 & b & 0.1635 & 0.0783 & 3.059 & 2.666 & 18.81 & 10.8745 & 0.3365 & 0.80 \\
2025W32\_2D\_0343\_4 & 3.5143 & 2460896.8555 &  & 0.1255 & 0.0797 & 3.222 & 2.711 & 17.04 & 9.5770 & 0.3116 & -0.22 \\
2025W33\_1B\_0642\_2 & 3.5149 & 2460901.9473 & b & 0.0080 & 0.0739 & 3.059 & 2.666 & 18.81 & 0.5289 & 0.0172 & 0.99 \\
2025W32\_2D\_0355\_1 & 3.6085 & 2460896.9169 &  & 0.1668 & 0.0708 & 3.220 & 2.711 & 17.06 & 12.7099 & 0.4309 & -0.15 \\
2025W33\_1B\_0642\_3 & 3.6096 & 2460901.9488 & b & -0.0132 & 0.0703 & 3.059 & 2.666 & 18.81 & -0.8747 & -0.0297 & 1.01 \\
2025W32\_2D\_0355\_2 & 3.7066 & 2460896.9184 &  & 0.1556 & 0.0671 & 3.220 & 2.711 & 17.06 & 11.8562 & 0.4265 & 0.12 \\
2025W33\_1B\_0642\_4 & 3.7069 & 2460901.9503 & b & 0.0291 & 0.0696 & 3.059 & 2.666 & 18.81 & 1.9381 & 0.0697 & 0.97 \\
2025W33\_1B\_0010\_3 & 3.8071 & 2460898.5514 & a & -0.1580 & 0.1548 & 3.168 & 2.695 & 17.65 & -11.5136 & -0.4329 & -2.09 \\
2025W32\_2D\_0355\_3 & 3.8096 & 2460896.9199 &  & 0.1520 & 0.0694 & 3.220 & 2.711 & 17.06 & 11.5754 & 0.4352 & 0.12 \\
2025W32\_2D\_0213\_4 & 3.8102 & 2460896.1734 & b & 0.9210 & 0.2429 & 3.244 & 2.718 & 16.79 & 71.5999 & 2.7140 & 0.93 \\
2025W33\_1B\_0010\_4 & 3.8408 & 2460898.5529 & b & -0.4225 & 0.1487 & 3.167 & 2.695 & 17.65 & -30.7858 & -1.1774 & -2.74 \\
2025W33\_1B\_0020\_1 & 3.8724 & 2460898.6147 &  & 0.0785 & 0.1479 & 3.166 & 2.694 & 17.67 & 5.7124 & 0.2224 & 0.15 \\
2025W33\_1B\_0020\_2 & 3.9059 & 2460898.6162 &  & 0.2967 & 0.1545 & 3.165 & 2.694 & 17.67 & 21.5821 & 0.8482 & 0.10 \\
2025W33\_1B\_0020\_3 & 3.9399 & 2460898.6177 &  & 0.2555 & 0.1502 & 3.165 & 2.694 & 17.68 & 18.5816 & 0.7447 & 0.23 \\
2025W33\_1B\_0020\_4 & 3.9742 & 2460898.6192 & b & 0.0697 & 0.1523 & 3.165 & 2.694 & 17.68 & 5.0716 & 0.2073 & 0.72 \\
2025W33\_1B\_0049\_1 & 4.0069 & 2460898.7511 &  & 0.2733 & 0.1558 & 3.161 & 2.693 & 17.72 & 19.8053 & 0.8169 & 0.07 \\
2025W33\_1B\_0049\_2 & 4.0418 & 2460898.7526 &  & 0.1826 & 0.1510 & 3.161 & 2.693 & 17.72 & 13.2307 & 0.5556 & 0.10 \\
2025W33\_1B\_0049\_3 & 4.0773 & 2460898.7541 &  & 0.2160 & 0.1577 & 3.161 & 2.693 & 17.72 & 15.6519 & 0.6700 & -0.39 \\
2025W33\_1B\_0049\_4 & 4.1127 & 2460898.7556 &  & 0.0984 & 0.1606 & 3.161 & 2.693 & 17.72 & 7.1327 & 0.3116 & -4.41 \\
2025W32\_2D\_0189\_1 & 4.1213 & 2460896.0344 & b & 0.3042 & 0.1611 & 3.248 & 2.720 & 16.74 & 23.7420 & 1.0372 & 0.32 \\
2025W31\_2A\_0219\_1 & 4.1342 & 2460889.1740 & b & 0.0495 & 0.1583 & 3.470 & 2.802 & 14.08 & 4.6787 & 0.2065 & 0.77 \\
2025W32\_2D\_0189\_2 & 4.1572 & 2460896.0359 & a & 0.1098 & 0.1627 & 3.248 & 2.720 & 16.74 & 8.5657 & 0.3819 & 0.57 \\
2025W31\_2A\_0219\_2 & 4.1703 & 2460889.1755 & b & 0.2781 & 0.1625 & 3.470 & 2.802 & 14.08 & 26.2830 & 1.1835 & 0.27 \\
2025W32\_2D\_0189\_3 & 4.1931 & 2460896.0374 &  & 1.9231 & 0.1885 & 3.248 & 2.720 & 16.74 & 150.0698 & 6.8255 & 0.04 \\
2025W31\_2A\_0219\_3 & 4.2063 & 2460889.1770 &  & 5.9340 & 0.2253 & 3.470 & 2.802 & 14.08 & 560.8176 & 25.5071 & 0.01 \\
2025W32\_2D\_0189\_4 & 4.2294 & 2460896.0389 &  & 35.0958 & 0.3847 & 3.248 & 2.720 & 16.74 & 2738.6005 & 125.8045 & 0.00 \\
2025W33\_1B\_0599\_1 & 4.2525 & 2460901.7420 &  & 53.6211 & 0.4486 & 3.066 & 2.667 & 18.74 & 3586.0773 & 168.0236 & -0.00 \\
2025W33\_1B\_0399\_1 & 4.2932 & 2460900.6541 &  & 16.2684 & 0.2982 & 3.100 & 2.676 & 18.38 & 1120.1553 & 53.5221 & 0.02 \\
2025W32\_2D\_0201\_1 & 4.3022 & 2460896.1047 &  & 8.8229 & 0.2521 & 3.246 & 2.719 & 16.76 & 687.2326 & 32.8367 & 0.01 \\
2025W33\_1B\_0399\_2 & 4.3300 & 2460900.6556 &  & 1.8853 & 0.1940 & 3.100 & 2.676 & 18.38 & 129.8094 & 6.3237 & 0.10 \\
2025W32\_2D\_0201\_2 & 4.3391 & 2460896.1062 & a & 0.5318 & 0.1862 & 3.246 & 2.719 & 16.77 & 41.4188 & 2.0177 & 0.29 \\
2025W33\_1B\_0399\_3 & 4.3668 & 2460900.6571 &  & 0.5401 & 0.1859 & 3.100 & 2.676 & 18.38 & 37.1844 & 1.8289 & 0.26 \\
2025W32\_2D\_0201\_3 & 4.3762 & 2460896.1077 & b & 0.4100 & 0.1807 & 3.246 & 2.719 & 16.77 & 31.9365 & 1.5859 & 0.36 \\
2025W33\_1B\_0399\_4 & 4.4042 & 2460900.6586 &  & 0.6458 & 0.1849 & 3.100 & 2.676 & 18.38 & 44.4580 & 2.2288 & 0.19 \\
2025W32\_2D\_0201\_4 & 4.4140 & 2460896.1092 & b & -0.2809 & 0.1862 & 3.246 & 2.719 & 16.77 & -21.8757 & -1.1072 & 55.34 \\
2025W32\_2D\_0480\_4 & 4.4273 & 2460897.5324 &  & 0.1900 & 0.2175 & 3.200 & 2.705 & 17.29 & 14.2339 & 0.7204 & -1.09 \\
2025W32\_1A\_0612\_1 & 4.4591 & 2460894.4706 & a & 0.1632 & 0.2409 & 3.299 & 2.736 & 16.15 & 13.2932 & 0.6856 & -3.15 \\
2025W32\_1A\_0612\_2 & 4.4913 & 2460894.4721 & c & 0.2695 & 0.2175 & 3.298 & 2.736 & 16.15 & 21.9529 & 1.1536 & -0.88 \\
2025W32\_1A\_0637\_1 & 4.5233 & 2460894.6067 &  & 0.0767 & 0.2141 & 3.294 & 2.735 & 16.20 & 6.2241 & 0.3301 & 0.24 \\
2025W32\_1A\_0637\_2 & 4.5564 & 2460894.6082 &  & 0.1235 & 0.2172 & 3.294 & 2.735 & 16.20 & 10.0205 & 0.5414 & -0.00 \\
2025W32\_1A\_0637\_3 & 4.5900 & 2460894.6097 & b & 0.0266 & 0.2223 & 3.294 & 2.735 & 16.20 & 2.1569 & 0.1176 & 0.60 \\
2025W32\_1A\_0637\_4 & 4.6237 & 2460894.6112 &  & 0.7050 & 0.2234 & 3.294 & 2.735 & 16.21 & 57.2192 & 3.1773 & 0.08 \\
2025W32\_2D\_0426\_4 & 4.7235 & 2460897.2641 & c & 0.3147 & 0.2634 & 3.209 & 2.707 & 17.19 & 23.7504 & 1.3792 & 0.02 \\
2025W32\_2D\_0017\_1 & 4.7389 & 2460895.0837 &  & 0.2891 & 0.2555 & 3.279 & 2.730 & 16.38 & 23.1563 & 1.3567 & -0.09 \\
2025W31\_2A\_0156\_1 & 4.7462 & 2460888.8341 & b & 0.0407 & 0.2621 & 3.481 & 2.807 & 13.94 & 3.8796 & 0.2273 & 0.90 \\
2025W32\_1A\_0662\_1 & 4.7716 & 2460894.7445 & b & 0.0042 & 0.2547 & 3.290 & 2.733 & 16.26 & 0.3357 & 0.0200 & 0.95 \\
2025W32\_2D\_0017\_2 & 4.7734 & 2460895.0852 &  & 0.2065 & 0.2546 & 3.279 & 2.730 & 16.38 & 16.5452 & 0.9866 & -0.26 \\
2025W31\_2A\_0156\_2 & 4.7805 & 2460888.8356 & b & -0.0536 & 0.2696 & 3.481 & 2.807 & 13.94 & -5.1183 & -0.3052 & 1.50 \\
2025W32\_1A\_0662\_2 & 4.8055 & 2460894.7460 & b & 0.0553 & 0.2594 & 3.290 & 2.733 & 16.26 & 4.4733 & 0.2691 & 0.32 \\
2025W32\_2D\_0017\_3 & 4.8081 & 2460895.0867 &  & 0.3201 & 0.2670 & 3.279 & 2.730 & 16.38 & 25.6390 & 1.5423 & -0.03 \\
2025W31\_2A\_0156\_3 & 4.8153 & 2460888.8371 &  & 0.4519 & 0.2628 & 3.481 & 2.807 & 13.94 & 43.1265 & 2.6168 & 0.09 \\
2025W32\_1A\_0662\_3 & 4.8408 & 2460894.7475 & a & -0.0689 & 0.2663 & 3.290 & 2.733 & 16.26 & -5.5716 & -0.3410 & 0.63 \\
2025W32\_2D\_0017\_4 & 4.8416 & 2460895.0882 & c & 0.5161 & 0.2797 & 3.279 & 2.730 & 16.39 & 41.3383 & 2.5301 & -0.01 \\
2025W31\_2A\_0156\_4 & 4.8501 & 2460888.8386 &  & 0.1717 & 0.2679 & 3.481 & 2.806 & 13.94 & 16.3823 & 1.0114 & -0.06 \\
2025W33\_1B\_0499\_1 & 4.8596 & 2460901.1996 & c & 0.4405 & 0.3001 & 3.083 & 2.672 & 18.56 & 29.8875 & 1.8451 & -0.07 \\
2025W32\_1A\_0662\_4 & 4.8765 & 2460894.7490 &  & 0.0851 & 0.2700 & 3.290 & 2.733 & 16.26 & 6.8800 & 0.4284 & -2.30 \\
2025W32\_2D\_0091\_2 & 4.9098 & 2460895.4945 &  & 0.1362 & 0.2893 & 3.266 & 2.725 & 16.54 & 10.7901 & 0.6777 & 6.99 \\
2025W32\_2D\_0091\_3 & 4.9463 & 2460895.4960 &  & 0.2143 & 0.3055 & 3.266 & 2.725 & 16.54 & 16.9733 & 1.0843 & -0.81 \\
2025W32\_2D\_0091\_4 & 4.9837 & 2460895.4975 &  & 0.1231 & 0.3297 & 3.265 & 2.725 & 16.54 & 9.7467 & 0.6306 & -3.60 \\
\enddata
\tablecomments{ObsID is the observation ID of \spherex~(each ID contains 6 frames from 6 arrays). $\lambda$ is the nominal band center of the pixel where \atlas\ center is located. Mid JD is the midpoint Julian date of the exposure. Flag indicates less reliable data (see text for detailed explanation). $F_\lambda$ and $dF_\lambda$ are the flux and uncertainty. $r_\mathrm{hel}$ and $r_\mathrm{obs}$ are the helio- and observer-centric distances to \atlas, respectively. $\alpha$ is the phase (Sun-target-observer) angle. $F_{\lambda, 1}$ is the reduced flux at $r_\mathrm{hel} = r_\mathrm{obs} = 1 \,\mathrm{au}$ assuming $F_\lambda \propto r_\mathrm{hel}^{-2} r_\mathrm{obs}^{-2}$. Refl is the reflectance normalized to 1 at $1.2\pm0.1 \um$. $F_\mathrm{field}$ is the fractional contamination from background objects relative to the aperture photometry measured on images \textit{before} subtraction of the simulated background within the aperture.}
\end{deluxetable}

\end{document}

%% file: author_list_3i-atlas_2025aug.tex
\author[0000-0002-9548-1526]{Carey~M.~Lisse}%
\affiliation{Johns Hopkins University, 3400 N Charles St, Baltimore, MD 21218, USA}%
\affiliation{Johns Hopkins University Applied Physics Laboratory, Laurel, MD 20723, USA}%
\email[show]{carey.lisse@jhuapl.edu}%

\author[0000-0002-2618-1124]{Yoonsoo~P.~Bach}%
\affiliation{Korea Astronomy and Space Science Institute (KASI), 776 Daedeok-daero, Yuseong-gu, Daejeon 34055, Republic of Korea}%
\email{ysbach93@gmail.com}%

\author[0000-0002-4650-8518]{Brendan~P.~Crill}%
\affiliation{Jet Propulsion Laboratory, California Institute of Technology, 4800 Oak Grove Drive, Pasadena, CA 91109, USA}%
\affiliation{Department of Physics, California Institute of Technology, 1200 E. California Boulevard, Pasadena, CA 91125, USA}%
\email{bcrill@jpl.nasa.gov}%

\author[0009-0003-8869-3651]{Phil~M.~Korngut}%
\affiliation{Department of Physics, California Institute of Technology, 1200 E. California Boulevard, Pasadena, CA 91125, USA}%
\email{pkorngut@caltech.edu}%

\author[0000-0002-7471-719X]{Ari~J.~Cukierman}%
\affiliation{Department of Physics, California Institute of Technology, 1200 E. California Boulevard, Pasadena, CA 91125, USA}%
\email{ajcukier@caltech.edu}%

\author[0000-0003-4607-9562]{Sean~A.~Bryan}%
\affiliation{School of Earth and Space Exploration, Arizona State University, 781 Terrace Mall, Tempe, AZ 85287 USA}%
\email{sean.a.bryan@asu.edu}%

\author[0000-0002-3892-0190]{Asantha~Cooray}%
\affiliation{Department of Physics \& Astronomy, University of California Irvine, Irvine CA 92697, USA}%
\email{acooray@uci.edu}%

\author[0009-0002-0098-6183]{C.~Darren~Dowell}%
\affiliation{Jet Propulsion Laboratory, California Institute of Technology, 4800 Oak Grove Drive, Pasadena, CA 91109, USA}%
\affiliation{Department of Physics, California Institute of Technology, 1200 E. California Boulevard, Pasadena, CA 91125, USA}%
\email{charles.d.dowell@jpl.nasa.gov}%

\author[0000-0003-4990-189X]{Michael~W.~Werner}%
\affiliation{Jet Propulsion Laboratory, California Institute of Technology, 4800 Oak Grove Drive, Pasadena, CA 91109, USA}%
\email{}%

\author[0000-0002-5599-4650]{Joseph~L.~Hora}%
\affiliation{Center for Astrophysics $|$ Harvard \& Smithsonian, Optical and Infrared Astronomy Division, Cambridge, MA 01238, USA}%
\email{jhora@cfa.harvard.edu}%

\author[0000-0003-4408-0463]{Zafar~Rustamkulov}%
\affiliation{IPAC, California Institute of Technology, MC 100-22, 1200 E California Blvd Pasadena, CA 91125, USA}%
\email{zafar@caltech.edu}%

\author[0000-0003-3119-2087]{Jeong-Eun Lee}%
\affiliation{Department of Physics and Astronomy, Seoul National University, 1 Gwanak-ro, Gwanak-gu, Seoul 08826, Republic of Korea}%
\email{lee.jeongeun@snu.ac.kr}%

\author[0000-0002-8244-4603]{Bumhoo~Lim}%
\affiliation{Department of Physics and Astronomy, Seoul National University, 1 Gwanak-ro, Gwanak-gu, Seoul 08826, Republic of Korea}%
\affiliation{SNU Astronomy Research Center, Department of Physics and Astronomy, Seoul National University, Gwanak-ro 1, Gwanak-gu, Seoul 08826, Republic of Korea}%
\email{bumhoo7@snu.ac.kr}%

\author[0000-0003-1156-9721]{Y.~R.~Fernandez}%
\affiliation{Department of Physics, University of Central Florida, Orlando, FL 32816-2385, USA}%
\email{Yanga.Fernandez@ucf.edu}%

\author[0000-0003-1841-2241]{Volker~Tolls}%
\affiliation{Center for Astrophysics $|$ Harvard \& Smithsonian, Optical and Infrared Astronomy Division, Cambridge, MA 01238, USA}%
\email{vtolls@cfa.harvard.edu}%

\author[]{W.~T.~Reach}%
\affiliation{Space Science Institute, 4765 Walnut Street, Suite B, Boulder, CO 80301, USA}%
\email{wreach@spacescience.org}%

\author[0000-0001-7432-2932]{O.~Dor\'{e}}%
\affiliation{Jet Propulsion Laboratory, California Institute of Technology, 4800 Oak Grove Drive, Pasadena, CA 91109, USA}%
\affiliation{Department of Physics, California Institute of Technology, 1200 E. California Boulevard, Pasadena, CA 91125, USA}%
\email{olivier.dore@caltech.edu }%

\author[0000-0001-8253-1451]{Michael~Zemcov}%
\affiliation{School of Physics and Astronomy, Rochester Institute of Technology, 1 Lomb Memorial Dr., Rochester, NY 14623, USA}%
\affiliation{Jet Propulsion Laboratory, California Institute of Technology, 4800 Oak Grove Drive, Pasadena, CA 91109, USA}%
\email{mbzsps@rit.edu}%

\author[0000-0002-5710-5212]{James~J.~Bock}%
\affiliation{Department of Physics, California Institute of Technology, 1200 E. California Boulevard, Pasadena, CA 91125, USA}%
\affiliation{Jet Propulsion Laboratory, California Institute of Technology, 4800 Oak Grove Drive, Pasadena, CA 91109, USA}%
\email{jjb@astro.caltech.edu}%

\author[0000-0002-5437-0504]{Yun-Ting~Cheng}%
\affiliation{Department of Physics, California Institute of Technology, 1200 E. California Boulevard, Pasadena, CA 91125, USA}%
\affiliation{Jet Propulsion Laboratory, California Institute of Technology, 4800 Oak Grove Drive, Pasadena, CA 91109, USA}%
\email{ycheng3@caltech.edu}%

\author[]{C.~Champagne}%
\affiliation{Northern Arizona University, Department of Astronomy and Planetary Science, Flagstaff, AZ, USA}%
\email{cc3776@nau.edu}%

\author[0009-0006-1028-1653]{Seungwon~Choi}%
\affiliation{Department of Physics and Astronomy, Seoul National University, 1 Gwanak-ro, Gwanak-gu, Seoul 08826, Republic of Korea}%
\affiliation{SNU Astronomy Research Center, Department of Physics and Astronomy, Seoul National University, Gwanak-ro 1, Gwanak-gu, Seoul 08826, Republic of Korea}%
\email{cygalbireo@snu.ac.kr}%

\author[]{M.~Connelley}%
\affiliation{University of Hawaii, 640 N. Aohoku Place, Hilo, HI 96720, USA}%
\email{msconnelley@gmail.com}%

\author[]{J.~P.~Emery}%
\affiliation{Northern Arizona University, Department of Astronomy and Planetary Science, Flagstaff, AZ, USA}%
\email{Joshua.Emery@nau.edu}%

\author[0000-0002-3745-2882]{Spencer~Everett}%
\affiliation{Department of Physics, California Institute of Technology, 1200 E. California Boulevard, Pasadena, CA 91125, USA}%
\email{severett@caltech.edu}%

\author[0000-0002-9382-9832]{Andreas~L.~Faisst}%
\affiliation{IPAC, California Institute of Technology, MC 100-22, 1200 E California Blvd Pasadena, CA 91125, USA}%
\email{afaisst@caltech.edu}%

\author[0000-0002-3291-4056]{Jooyeon~Geem}%
\affiliation{Asteroid Engineering Laboratory, Lule\aa University of Technology, Box 848, Kiruna, 98128, Sweden}%
\email{ksky0422@gmail.com}%

\author[0000-0001-5812-1903]{Howard~Hui}%
\affiliation{Department of Physics, California Institute of Technology, 1200 E. California Boulevard, Pasadena, CA 91125, USA}%
\email{hhui@caltech.edu}%

\author[0000-0002-7332-2479]{Masateru~Ishiguro}%
\affiliation{Department of Physics and Astronomy, Seoul National University, 1 Gwanak-ro, Gwanak-gu, Seoul 08826, Republic of Korea}%
\affiliation{SNU Astronomy Research Center, Department of Physics and Astronomy, Seoul National University, Gwanak-ro 1, Gwanak-gu, Seoul 08826, Republic of Korea}%
\email{ishigrmt@gmail.com}%

\author[0000-0002-0460-7550]{Sunho~Jin}%
\affiliation{Korea Astronomy and Space Science Institute (KASI), 776 Daedeok-daero, Yuseong-gu, Daejeon 34055, Republic of Korea}%
\email{jsh854@naver.com}%

\author[0009-0004-9591-8646]{Hangbin~Jo}%
\affiliation{Department of Physics and Astronomy, Seoul National University, 1 Gwanak-ro, Gwanak-gu, Seoul 08826, Republic of Korea}%
\affiliation{SNU Astronomy Research Center, Department of Physics and Astronomy, Seoul National University, Gwanak-ro 1, Gwanak-gu, Seoul 08826, Republic of Korea}%
\email{hangbin9@naver.com}%

\author[0000-0003-2831-0513]{Max~Mahlke}%
\affiliation{Universit'{e} Marie et Louis Pasteur, CNRS, Institut UTINAM (UMR 6213), '{e}quipe Astro, F-25000 Besan,{c}on, France}%
\email{ max.mahlke@rwth{-}aachen.de}%

\author[0000-0001-5382-6138]{Daniel~C.~Masters}%
\affiliation{IPAC, California Institute of Technology, MC 100-22, 1200 E California Blvd Pasadena, CA 91125, USA}%
\email{dmasters@ipac.caltech.edu}%

\author[0000-0002-6025-0680]{Gary~J.~Melnick}%
\affiliation{Center for Astrophysics $|$ Harvard \& Smithsonian, Optical and Infrared Astronomy Division, Cambridge, MA 01238, USA}%
\email{gmelnick@cfa.harvard.edu}%

\author[0000-0001-9368-3186]{Chi~H.~Nguyen}%
\affiliation{Department of Physics, California Institute of Technology, 1200 E. California Boulevard, Pasadena, CA 91125, USA}%
\email{chnguyen@caltech.edu}%

\author[0000-0002-5158-243X]{Roberta~Paladini}%
\affiliation{IPAC, California Institute of Technology, MC 100-22, 1200 E California Blvd Pasadena, CA 91125, USA}%
\email{paladini@ipac.caltech.edu}%

\author[]{M.~L.~Sitko}%
\affiliation{Space Science Institute, 4765 Walnut Street, Suite B, Boulder, CO 80301, USA}%
\email{sitko@SpaceScience.org}%

\author[0000-0003-3078-2763]{Yujin~Yang}%
\affiliation{Korea Astronomy and Space Science Institute (KASI), 776 Daedeok-daero, Yuseong-gu, Daejeon 34055, Republic of Korea}%
\email{yyang@kasi.re.kr}%